\documentclass[sigconf,10pt]{acmart}

\usepackage{epsfig,endnotes}
\usepackage{textcomp} 
\usepackage[english]{babel}
\usepackage{times}
\usepackage{microtype}
\usepackage{amsfonts,amsmath}
\usepackage{balance}
\usepackage{graphicx}
\usepackage{multirow}
\usepackage{paralist}
\usepackage[font=small,labelfont=bf]{caption}
\usepackage{subcaption}
\usepackage{xspace}
\usepackage{booktabs}
\usepackage{listings}
\usepackage{url}

\usepackage{breakurl}

\usepackage{hyperref}
\usepackage{listings}
\usepackage{enumitem}
\usepackage{xcolor}
\usepackage{siunitx}
\usepackage{algorithm}
\floatname{algorithm}{Component}
\usepackage[noend]{algpseudocode}
\usepackage{array}
\usepackage{balance}
\usepackage[breakable, theorems, skins]{tcolorbox}
\usepackage{float}
\usepackage{mathtools}

\usepackage{csquotes}

\usepackage{graphicx}

\renewcommand\footnotetextcopyrightpermission[1]{} 
\setcopyright{none}

\newcommand{\name}{\textsc{FB}\xspace}
\newcommand{\app}{\textsc{FBA}\xspace}

\newcommand{\restr}[1]{\textit{\textbf{C#1}\xspace}}
\newcommand{\q}[1]{\textit{\textbf{Q#1}\xspace}}

\newcommand{\remove}[1]{}
\newcommand{\myitem}[1]{\vspace*{0.02in}\noindent\textbf{#1}}

\newcommand{\cref}[1]{Chapter~\ref{#1}}

\newcommand{\fref}[1]{Fig.~\ref{#1}}

\newcommand{\first}{\emph{(i)}\xspace}
\newcommand{\second}{\emph{(ii)}\xspace}
\newcommand{\third}{\emph{(iii)}\xspace}
\newcommand{\fourth}{\emph{(iv)}\xspace}

\newcommand{\ie}{i.e., \@}
\newcommand{\eg}{e.g., \@}

\newcommand*\circled[1]{\tikz[baseline=(char.base)]{
	\node[shape=circle,draw,inner sep=1pt] (char) {#1};}}

\begin{document}
\date{}

\title[]{\LARGE \bf \name: A Flexible Buffer Management Scheme \\ for Data Center Switches}

\author{\Large Maria Apostolaki}
\affiliation{\Large ETH Zurich}
\author{\Large Vamsi Addanki}
\affiliation{\Large ETH Zurich}
\author{\Large Manya Ghobadi}
\affiliation{\Large MIT}
\author{\Large Laurent Vanbever}
\affiliation{\Large ETH Zurich}

\renewcommand{\shortauthors}{Apostolaki et al.}

\begin{abstract}

Today, network devices share buffer across priority queues to avoid drops during transient congestion.
While cost-effective most of the time, this sharing can cause undesired interference among seemingly independent traffic.
As a result, low-priority traffic can cause increased packet loss to high-priority traffic.
Similarly, long flows can prevent the buffer from absorbing incoming bursts even if they do not share the same queue.
The cause of this perhaps unintuitive outcome is that today's buffer sharing techniques are unable to guarantee isolation across (priority) queues without statically allocating buffer space.
To address this issue, we designed \name, a novel buffer sharing scheme that offers strict isolation guarantees to high-priority traffic without sacrificing link utilizations.
Thus, \name outperforms conventional buffer sharing algorithms in absorbing bursts while achieving on-par throughput.
We show that \name is practical and runs at line-rate on existing hardware (Barefoot Tofino).
Significantly, \name's operations can be approximated in non-programmable devices.

\end{abstract}

\maketitle
\pagestyle{empty}

\section{Introduction}

To reduce cost and maximize utilization, network devices often rely on a
shared buffer chip whose allocation across queues is dynamically adjusted
by a buffer management algorithm~\cite{das2012broadcom,lect12,mathisbuffer}.
The most commonly-used buffer management algorithm today is
Dynamic Thresholds (DT)~\cite{choudhury1998dynamic,broadcom_patent,das2012broadcom,ciscoDT,heunderstanding}.
DT dynamically allocates buffer per queue proportionally to the still-unoccupied buffer space.
As a result, the more the queues are using the buffer, the less buffer each of them is allowed to occupy.

Despite its wide deployment, DT does not meet the requirements of today's multi-tenant data-center environments 
for three key reasons.
First, DT cannot reliably absorb bursts, which are of paramount importance for application performance~\cite{chen2009understanding,phanishayee2008measurement}.
Second, DT is unable to offer any isolation guarantee, meaning that the performance of traffic (even of high priority)
is dependent on the instantaneous load on each device it traverses. 
Finally, DT is unable to react to abrupt changes in the traffic demand, as it keeps the buffer highly utilized (to improve throughput), even if this brings little benefit. 
Worse yet, as we shall show, more sophisticated approaches \eg Cisco IB~\cite{cisco9000} inherit DT's limitations.
To compensate for these limitations, data-center operators often statically allocate part of the buffer space to queues, effectively wasting precious buffer space that could be put to better use (\eg to absorb bursts).

While Congestion Control (CC)
algorithms and scheduling techniques can alleviate the shortcomings of DT, they are unable to address them fully.
Indeed, CC could decrease the buffer utilization, indirectly leaving more space for bursts, while scheduling could allow preferential treatment of certain priority queues across those sharing a single port.
Yet, each of these techniques senses and controls distinct network variables.
First, CC can only sense per-flow performance (\eg loss or delay) but is oblivious to the state of the shared buffer (buffer pressure~\cite{alizadeh2011data}) and the relative priority across competing flows. Worse yet, CC controls the rate of a given flow but cannot affect the rate at which other flows are sending.
Thus, CC cannot resolve buffer conflicts across flows sharing the same device.
Second, scheduling can only sense the per-queue occupancy and control the transmission (dequeue) of packets via a particular port \emph{after and only if} they have been enqueued. As a result,
scheduling cannot resolve buffer conflicts across queues not sharing the same port.

To address DT's shortcomings without statically allocating buffer, we design \name.
 \name is a novel buffer
management algorithm that manages the buffer as a multi-dimensional entity. Unlike DT,
\name allocates buffer by sensing not only the absolute buffer occupancy but also its content and temporal characteristics.
In particular, \name \emph {(i)} guards the distribution of traffic across multiple dimensions (\eg per priority), preventing sets of queues from monopolizing the buffer; and \emph{(ii)} favors queues that free-up their used buffer faster,
making the buffer available to more traffic.
As a result, \name can keep the buffer usage under control and offer \emph{provable} performance guarantees \emph{without} resorting to static allocations.

Despite its benefits, \name is deployable today.
We implemented \name on a programmable device (Barefoot Tofino Wedge 100BF-32X~\cite{tofino}).
We also describe how to approximate \name's behavior by periodically reconfiguring DT with reduced-yet-significant benefits.

We show that \name outperforms alternative buffer
management techniques even when they are combined with DCTCP. 
Indeed, \name (with TCP) improves burst absorption (measured as Query Completion Time) compared to Dynamic Thresholds with DCTCP
 by 10\% (53\%) when the network utilization is 40\% (90\%).  
Moreover, \name reduces the Flow Completion Time (FCT) 99-th percentile of DT with DCTCP by 38\% even when the network utilization is 20\%. 
Importantly, \name achieves on-par throughput compared to other techniques. 
While \name's benefits increase with contention of the buffer, it never deteriorates performance, even at low utilization or in the absence of high-priority or bursty traffic.

\noindent Our main contributions include:

\begin{itemize}
\item The first analysis of DT (the most widely-used buffer management algorithm today) in a multi-queue setting. Our analysis reveals DT's inefficiencies both experimentally (\S\ref{sec:evaluation}) and analytically (\S\ref{sec:background}).

\item A novel approach for buffer management that can provide performance guarantees without statically allocating buffer (\S\ref{sec:design}).

\item A novel hardware design and implementation of \name on a Barefoot Tofino switch~\cite{tofino} that demonstrates its practicality in today's hardware (\S\ref{ssec:pisa}), together with an approximation of its behavior that extends its deployability to more-commonly-used devices (\S\ref{ssec:pca}).

\item A comprehensive evaluation demonstrating that \name outperforms state-of-the-art buffer management algorithms even when combined with DCTCP (\S\ref{sec:evaluation}).

\end{itemize}

\section{Background \& Motivation}
\label{sec:background}
In this section, we first describe our model, namely the network device architecture we consider (\S\ref{ssec:dt1}).
Next, we explain how \emph{Dynamic Thresholds} (DT), the most commonly-used buffer management algorithm, works (\S\ref{ssec:dt2}).
Finally, we reveal DT's core inefficiencies (\S\ref{ssec:dt3}).

\vspace{-1mm}

\subsection{\textbf{Switch Model}} \label{ssec:dt1}
\begin{figure}[t]
\centering
\includegraphics[width=0.95\linewidth]{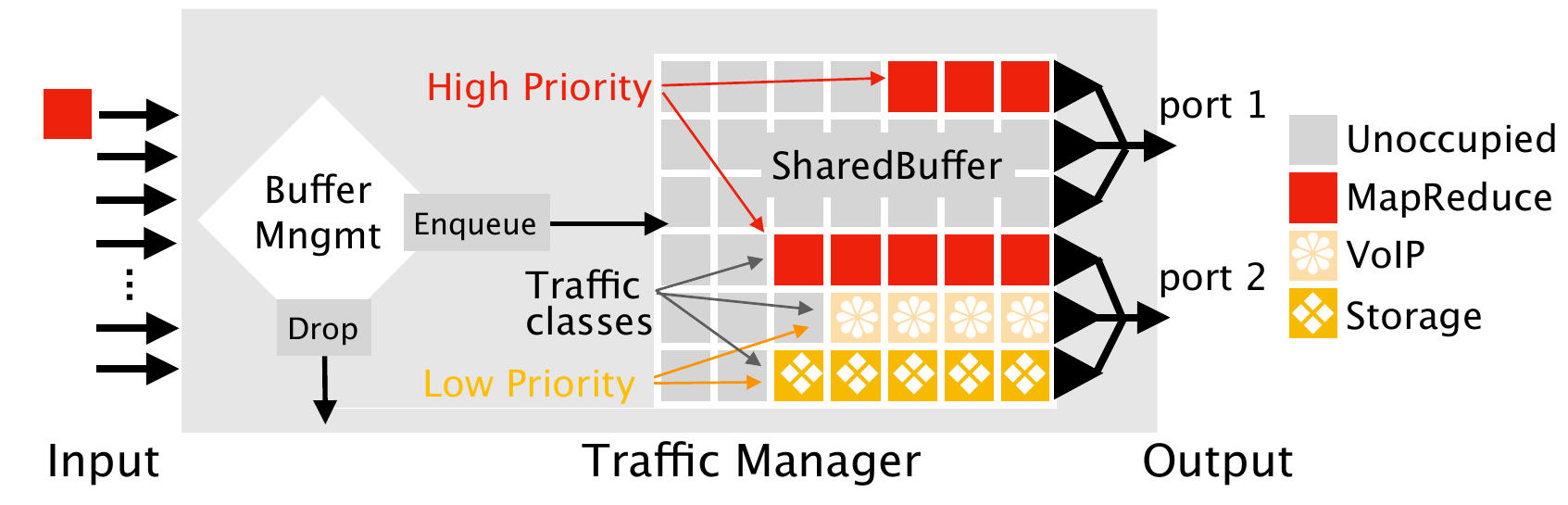}
  \caption[]{Traffic is grouped into classes (\eg MapReduce, VoIP) and each class is of high or low priority.
  The Traffic Manager stores an incoming packet in the shared buffer iff the length of the corresponding queue is below
  the threshold that it calculates according to the buffer management algorithm.  }%
\vspace{1mm}
\label{fig:bufIntro}
\end{figure}

\fref{fig:bufIntro} shows a simplified output-queued shared-memory packet switch.\footnote{We describe the mapping of our model to RMT architecture in~\S\ref{sec:hwdesign}.}
The switch implements a fixed or programmable logic which maps each packet to a particular queue of a port.
The switch stores incoming packets in its buffer for future transmission. The switch cannot store all incoming packets, as
 the space in the buffer is limited.
A mechanism in the switch, namely the Traffic Manager (TM)\footnote{TM's architecture is the same for both fixed-function and reconfigurable switches~\cite{sharma2020programmable}, including Barefoot's Tofino and Broadcom's Trident series.} determines whether to store or to drop incoming packets. To that end, the TM compares the queue's length with a threshold that it calculates according to a buffer management algorithm, \eg DT.

We assume that the operator groups traffic into \textbf{classes}.
Each class exclusively uses a single queue at each port to achieve cross-class \emph{delay isolation}~\cite{qos}. For instance, in Fig.~\ref{fig:bufIntro}, Storage, VoIP and MapReduce belong to distinct traffic classes.

We also assume that each traffic class is of \textbf{high or low priority}.
Distinguishing classes to high and low priority facilitates prioritizing of certain classes over others in times of high load.
This prioritization concerns the use of the shared buffer and does not affects scheduling.
The operator can configure multiple low-priority classes and multiple high-priority classes.
In a cloud environment, traffic that is subject to Service Level Agreements (SLAs) would be high-priority.
In Fig.~\ref{fig:bufIntro} the MapReduce class is of high priority, while \emph{all} other classes are of low priority.

\subsection{\textbf{DT's workings}} \label{ssec:dt2}
We now describe Dynamic Thresholds (DT), the most common buffer management algorithm in today's devices~\cite{broadcom_patent,ciscoDT,heunderstanding,miao2019buffer,apostolaki2019fab}.
DT~\cite{choudhury1998dynamic} dynamically adapts the instantaneous maximum length of each queue, namely its threshold $T_c^i(t)$ according to the remaining
buffer and a configurable parameter $\alpha$, as we see in Eq.\eqref{eq:dt}.
DT's per-queue threshold is: \first
directly proportional to the remaining buffer ($B-Q(t)$) \ie the less unoccupied buffer there
is, the less a queue can grow; and \second directly proportional to a
parameter $\alpha$ often configured per class\footnote{While $\alpha$ can be configured per queue,
it is often configured per class.}: the
higher the $\alpha$, the more the queue can grow.

\begin{equation}
T_c^i(t) = \alpha_c^i \cdot (B-Q(t))
\label{eq:dt}
\end{equation}
\begin{equation*}
\begin{aligned}
\vspace{-1mm} &\text{$T_c^i(t)$: Queue threshold of class $c$ in port $i$} \\
\vspace{-1mm} &\text{$\alpha_c^i$: $\alpha$ parameter of class $c$ in port $i$} \\
\vspace{-1mm} &B: \text{Total buffer space \footnotemark  \qquad \qquad \qquad \qquad \qquad} \\
\vspace{-1mm} &Q(t): \text{ Total buffer occupancy at time $t$}
\end{aligned}
\end{equation*}

\footnotetext{For simplicity, we assume a single buffer-chip per device.}

The $\alpha$ parameter of a queue impacts its maximum length and its relative length with respect to the other queues. Thus, an operator is likely to set higher $\alpha$ values for high-priority traffic classes compared to low-priority ones\footnote{We use 'priorities' to categorize traffic classes; it is not related to scheduling.}.
Despite its importance, there is no systematic way to configure $\alpha$, meaning
different vendors and operators reportedly use different $\alpha$ values. For instance, Yahoo uses $\alpha=8$~\cite{heunderstanding} while Cisco $\alpha=14$~\cite{ciscoDT} and Arista $\alpha=1$.

\myitem{}The buffer alternates between steady and transient state:\\
\myitem{Steady-state} is the state during which all queues sharing the buffer are shorter or equal to the threshold that DT calculates.\\
\myitem{Transient-state} is the state during which at least one queue in the buffer is longer than its threshold.\\

\begin{figure}[t]
\centering
\includegraphics[width=0.98\linewidth]{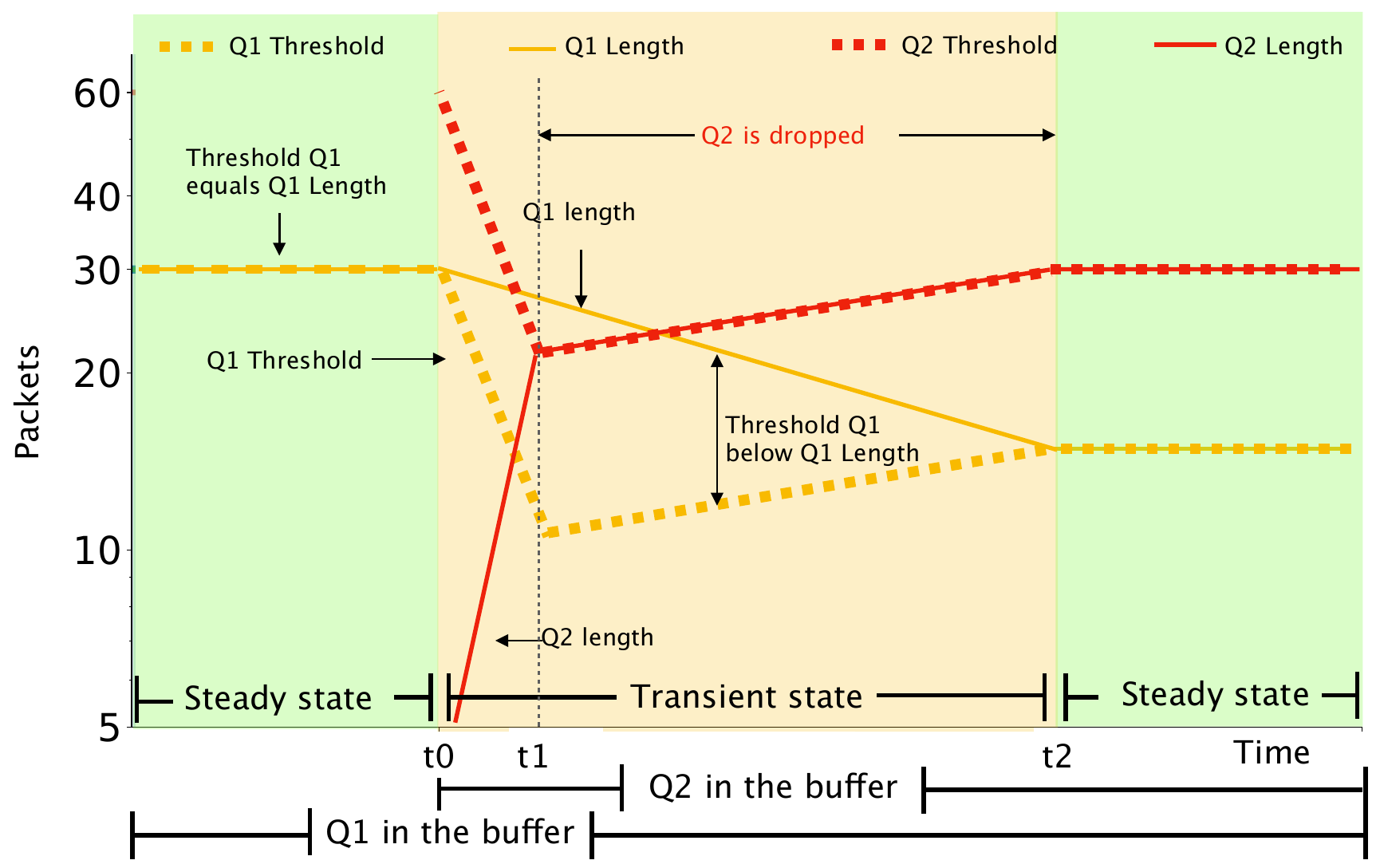}
  \caption[]{At $t0$ an incoming burst (Q2) rapidly changes the buffer occupancy.
  In the transient state ($t0-t2$), the threshold of Q1 is lower than its length.
  Thus all its incoming packets are dropped to free buffer for Q2.
  Still, Q2 experiences drops before reaching its fair steady-state allocation ($t1-t2$).}
  \label{fig:transientState}
\vspace{9pt}
\end{figure}

\myitem{DT in action.}
To understand how the buffer alternates between steady and transient state, we walk through
an example scenario.
Consider a switch with a $60$-packet buffer shared across multiple queues mapped to two distinct traffic classes; one of high and one of low priority.
\fref{fig:transientState} illustrates the evolution of the length of two queues and their thresholds over time, illustrated in solid and dotted lines respectively.
One queue, say Q1 belongs to the low-priority class and is colored in yellow.
The other queue, say Q2 belongs to the high-priority class, is colored in red.
The operator has configured $\alpha=1$ for the class of Q1 and $\alpha=2$ for the class of Q2.

Before time $t0$, the only non-empty queue in the buffer is the low-priority yellow class: Q1.
During this time, the buffer is in steady state, meaning all queues' length is lower or equal to DT's thresholds.
Indeed, Q1's length is $30$ packets, the remaining buffer space is equal to $30$ packets; thus from Eq.~\eqref{eq:dt} Q1's threshold is $1 \times 30 = 30$.

At time $t0$, a burst of packets belonging to the high-priority red class arrives at the switch.
The burst causes Q2's length to increase in the time frame ($t0-t2$) (solid red line).

At time $t1$, Q2's length increase is inhibited: Q2 continues to grow but at a lower rate as its length starts being controlled by Q2's threshold (red dashed line), which DT calculates.
Q2's threshold decreases due to Q2's growth in the buffer, which reduces the overall remaining buffer.
Thus, during the time frame ($t1-t2$), some of the packets mapped to Q2 are dropped.
In the time frame ($t0-t1$), the reduction of the remaining buffer also causes Q1's threshold to decrease.
Notably, Q1's threshold decreases at a rate higher than its length does. In the time frame ($t0-t2$), the buffer is in transient state, as Q1's length is higher than its threshold.

At $t2$, the buffer reaches steady state again. This time, the remaining buffer is $15$ packets, Q1 occupies $15$ packets, resulting from Q1's threshold $T_{q1} =1 \times 15$ packets, while Q2 occupies $30$, as $T_{q2}=2 \times 15$ .

\myitem{To sum up}, the high-priority burst was dropped \emph{before} the buffer had reached steady state.
Importantly, these drops could have been avoided if
\emph{(i)} there was more available buffer when the burst arrived (steady-state allocation); or \emph{(ii)} the buffer could have been emptied faster to make room for the burst (transient-state allocation).

\subsection{DT's inefficiencies}\label{ssec:dt3}

Having explained the importance of steady and transient-state allocation in the buffer management's behavior, we now
explain why DT is fundamentally unable to \emph{control} any of the two.
In particular, we show analytically and using intuitive examples that DT
blindly maximizes the buffer utilization at the expense of predictable allocations.
As a result, DT cannot offer any minimum buffer guarantee nor any burst-tolerance guarantee.
The former is only affected by DT's steady-state allocation, while the latter by both steady- and transient-state allocation, as we explained in~\S\ref{ssec:dt2}

\begin{figure}[hb]
\centering
\includegraphics[width=0.9\linewidth]{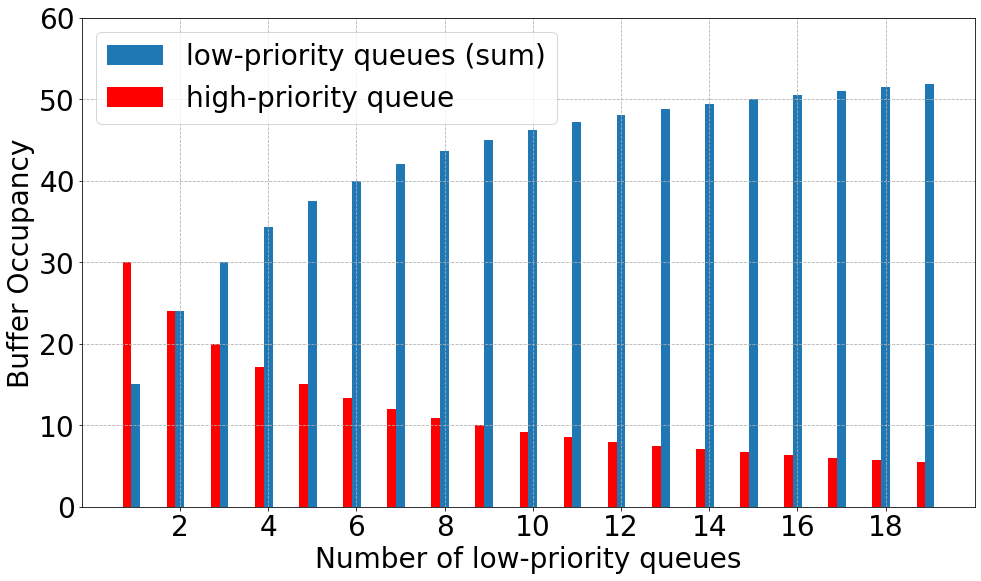}
  \caption[]{High-priority traffic can use only a fraction of the buffer because DT allocates unrestrictedly
 more buffer to low-priority queues as their number increases (x-axis).}%
\label{fig:dtanalysis}
\vspace{-3pt}
\end{figure}

\begin{figure*}
 \centering

 \begin{subfigure}[t]{0.24\textwidth}
  \centering
  \includegraphics[width=\textwidth]{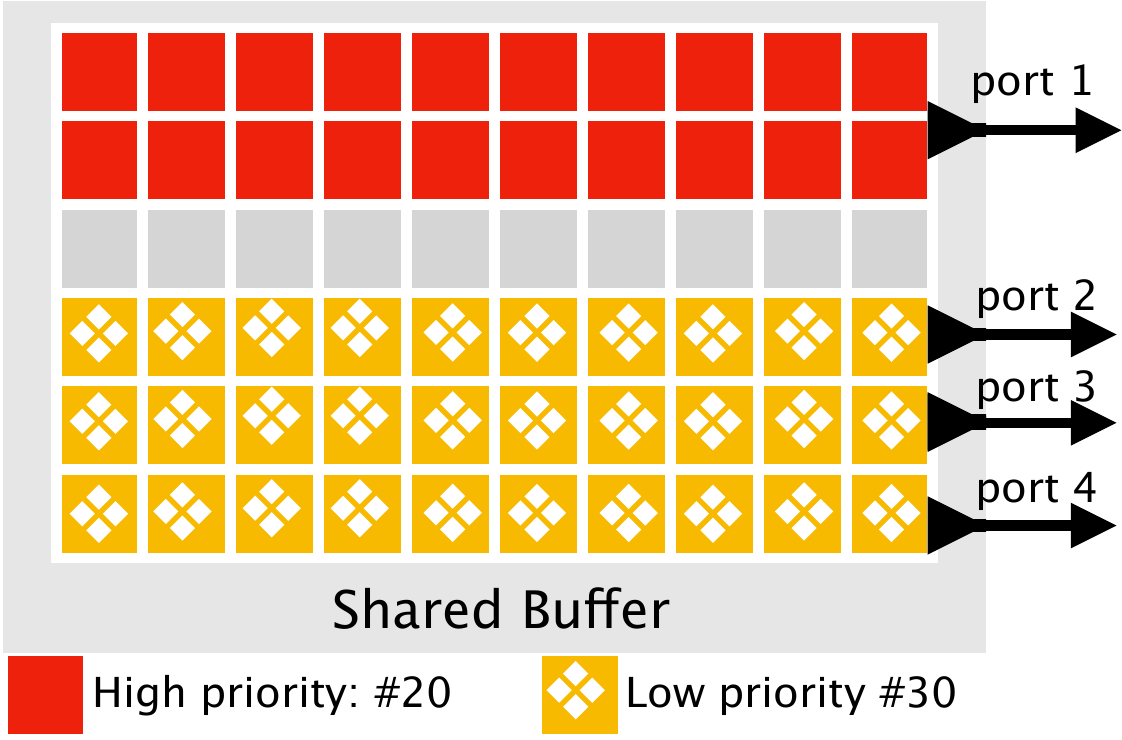}
  \caption[]{DT allocates more buffer to low-priority queues than the high one.}%
  {{\small}}
  \label{fig:period3}
 \end{subfigure}
  \hfill
  \begin{subfigure}[t]{0.35\textwidth}
  \centering
  \includegraphics[width=\textwidth]{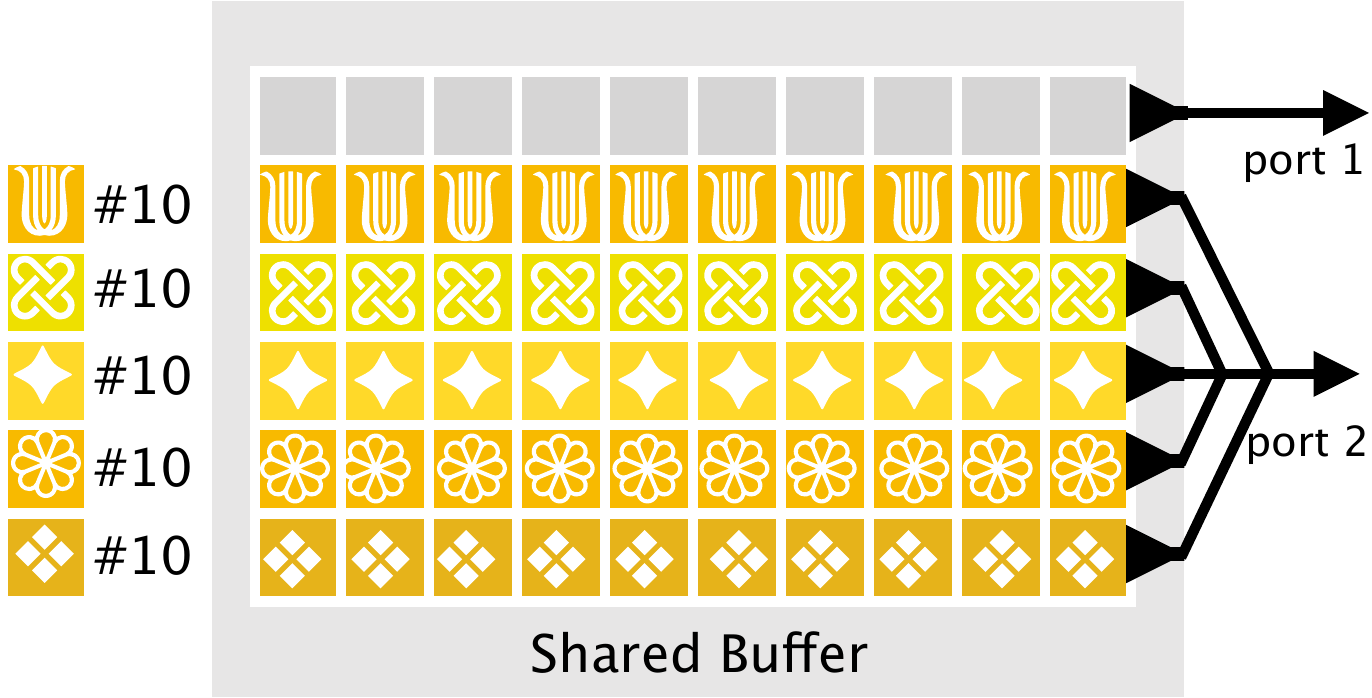}
  \caption[]{DT allows queues sharing the same port to grow as much as it would if they were using different ones.}%
  {{\small }}
  \label{fig:example4}
 \end{subfigure}
  \hfill
  \begin{subfigure}[t]{0.35\textwidth}
  \centering
  \includegraphics[width=\textwidth]{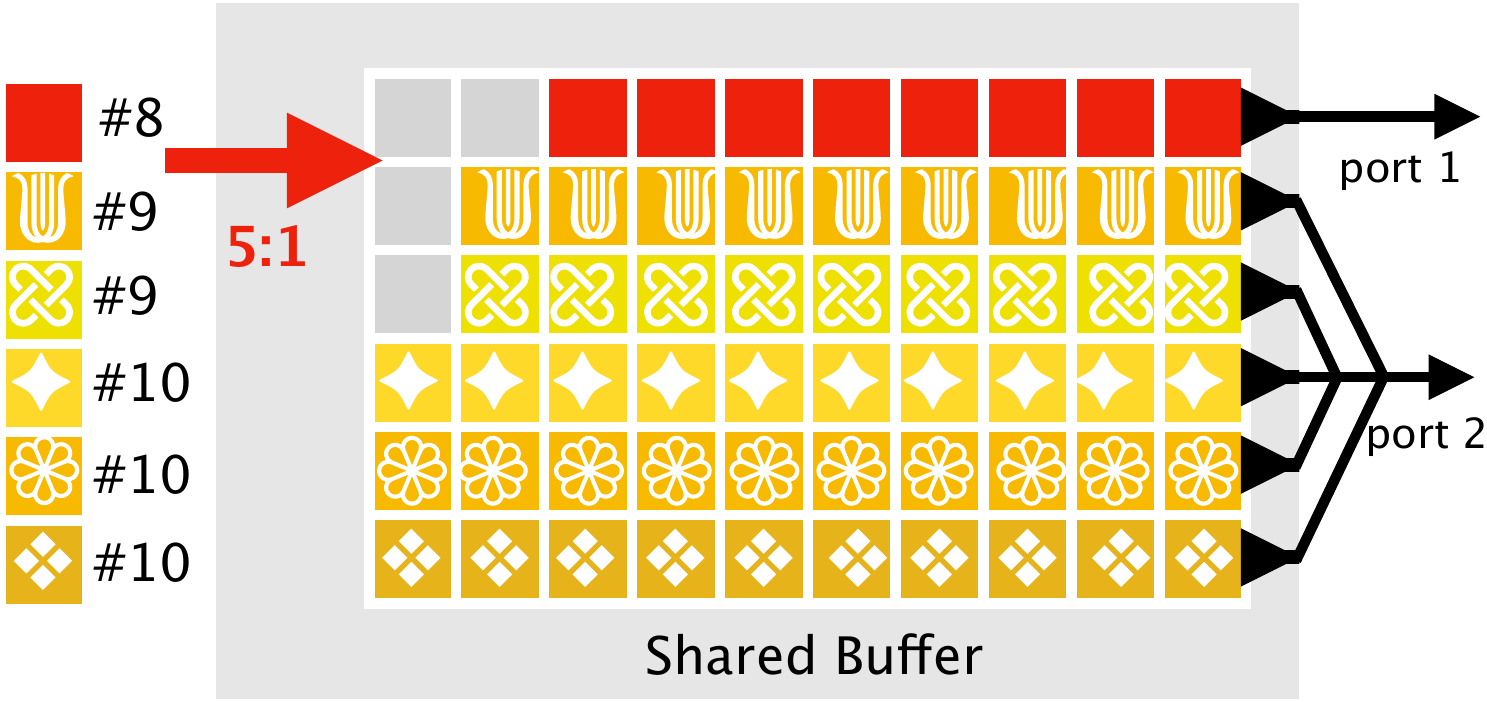}
  \caption[]{A burst of the red class is dropped due to DT's transient state unpredictability.}%
  {{\small}}
  \label{fig:example5}
 \end{subfigure}
\caption[]{DT allows high buffer occupancy with low aggregate dequeue rate, causing poor burst-tolerance capabilities.}
\label{fig:prob2}
\end{figure*}

\myitem{DT offers no minimum buffer guarantee.}
DT enforces the precedence of a queue or class over the others via a \emph{static} parameter ($\alpha$).
Yet, $\alpha$ offers no guarantee as the actual per-queue threshold depends on the overall remaining buffer (Eq.~\eqref{eq:dt}), which
 can reach arbitrarily and uncontrollably low values, even in the steady state.
 \footnote{An operator can statically allocate space for each queue; in this case, though, the buffer space will be wasted unless used by the corresponding queue.}

To better understand this limitation, consider the switch shown in~\fref{fig:period3} with a $60$-packet buffer shared across four non-empty queues. Each queue is mapped to a distinct port. The queue of port $1$ belongs to a high-priority class and is colored in red; The three other queues mapped to port $2-4$ belong to a low-priority class and are colored in yellow.

The operator intended to preferentially treat the high-priority class. Thus, she configures $\alpha=2$ for the high-priority and $\alpha=1$ for low-priority. Indeed, this configuration allows the red queue to occupy a large portion of the buffer (up to 40 packets) and $2x$ more buffer than any yellow queue.

In the practice though, the red queue's threshold is unpredictable and can get very low.
Indeed, in the illustrated instance (\fref{fig:period3}), the red queue's threshold (and length) is $20$ packets as the remaining is $10$ and $\alpha=2$. The three other queues collectively occupy $30$ packets, namely more than the high-priority red queue.

In fact, the buffer available to a high-priority queue can even approach zero. As an illustration, Fig.~\ref{fig:dtanalysis} shows the buffer that a high-priority queue occupies compared to the aggregate buffer the low-priority queues occupy as a function of the number of low-priority congested queues\footnote{As a reminder we use high-priority as a term to define the importance of a traffic class and is not related to scheduling.}. This insight is experimentally supported by~\cite{van2019empirical}, in which authors observed the behavior of programmable switches of different vendors.
While the operator could potentially configure $\alpha$ values to achieve a desirable buffer distribution \emph{given the number of congested queues in all ports}, the latter cannot be accurately predicted.

To sum up, when DT is used, the buffer occupancy and thus the remaining buffer depends on the number of congested queues, and that, independently of the priority or class they belong to.
Indeed, as we observe in Eq.~\ref{eq:BufferOcupancy}, the buffer occupancy $Q(t)$ increases with the number of congested queues $N$ in the numerator \ie $\sum_{N}\alpha_c^i$.

\begin{equation}
\label{eq:BufferOcupancy}
Q(t) = \frac{B\cdot  \displaystyle\sum_{N}\alpha_c^i} {1 + \displaystyle\sum_{N}\alpha_c^i }
\end{equation}

\begin{equation*}
\begin{aligned}
\vspace{-1mm} &\text{$Q(t)$:Buffer occupied} \\
\vspace{-1mm} &\text{$N$: set of queues (i,c) that are congested} \\
\vspace{-1mm} &\\
\end{aligned}
\end{equation*}

\myitem{DT offers no burst-tolerance guarantees.}
In addition to the unpredictability of its steady-state allocation, DT's transient-state allocation is uncontrollable.
This is particularly problematic when it comes to burst absorption.
The main reason for this limitation is that DT perceives buffer space as a scalar quantity ignoring its
expected occupancy over time.

To better understand this limitation we use an intuitive example shown in \fref{fig:example4} and \fref{fig:example5}. The two figures illustrate the same 60-packet buffer before and after the arrival of a burst.

In \fref{fig:example4}, the buffer is shared across five non-empty queues of low-priority classes (all in yellow colors).
Notably, all queues are mapped to a single port (port 2).
DT allows each queue to occupy $10$ packets, as they are configured with $\alpha=1$ and the remaining buffer space in the steady-state is $10$.

In \fref{fig:example5}, a high-priority 5:1 incast occurs at port 1, meaning $5$ incoming ports simultaneously send to port 1.
Due to DT's prior allocations, though, the buffer cannot keep up with the incoming traffic. Concretely, the buffer has not enough unoccupied space, nor can it be emptied fast enough to make room for the burst.
As a result, the high-priority burst (caused by the incast) starts to experience drops while it only occupies $8$ packets in the buffer, that is $9$ packets less than the steady-state allocation of the corresponding queue,
which would be $\sim17$ packets.
In other words, the high-priority traffic class is experiencing drops in the transient state,
which could have been avoided if the buffer could reach steady state faster.
The reason of this slowdown is that the $5$ low-priority queues share the dequeue rate of a single port.
Notably, DT has no way to distinguish between $5$ queues coexisting in a single port and $5$ queues, each attached to a separate port.

\begin{figure*}[t]
 \centering
  \begin{subfigure}[t]{0.24\textwidth}
  \centering
  \includegraphics[width=\textwidth]{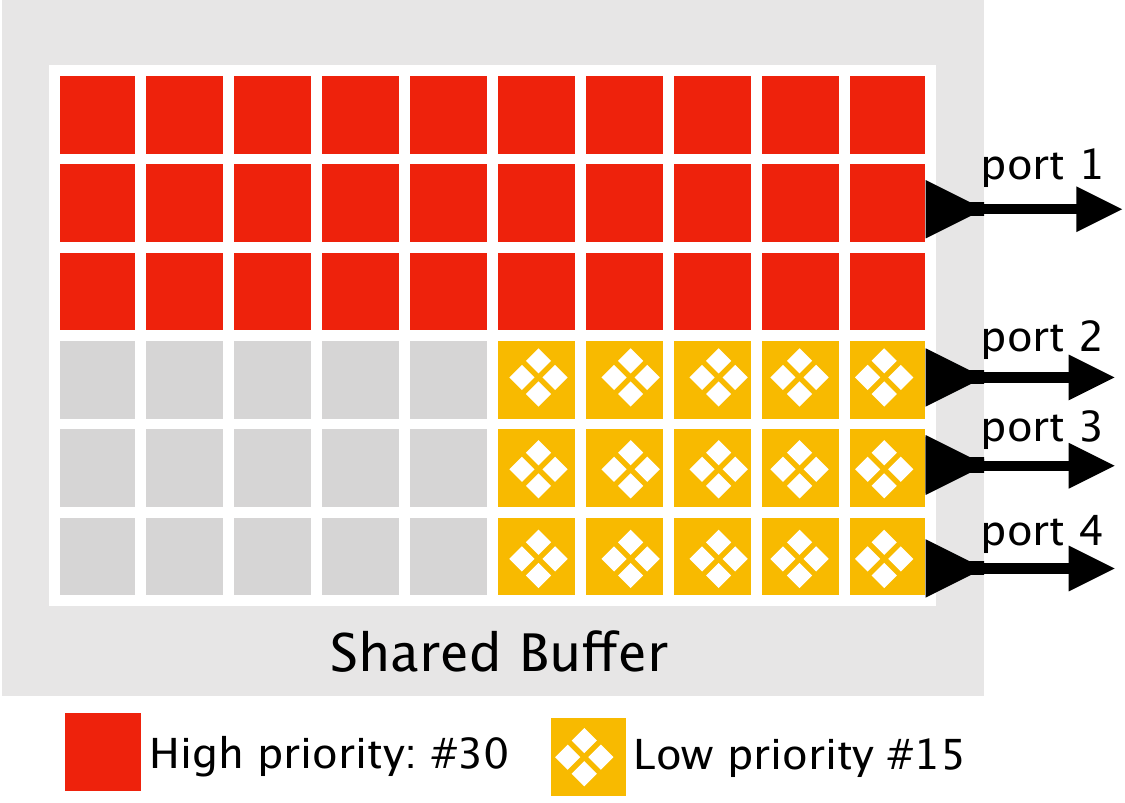}
  \caption[]{\name guarantees the high-priority queue a minimum allocation.}  
  {{\small}}
  \label{fig:example3}
 \end{subfigure}
 \hfill
 \begin{subfigure}[t]{0.32\textwidth}
  \centering
  \includegraphics[width=\textwidth]{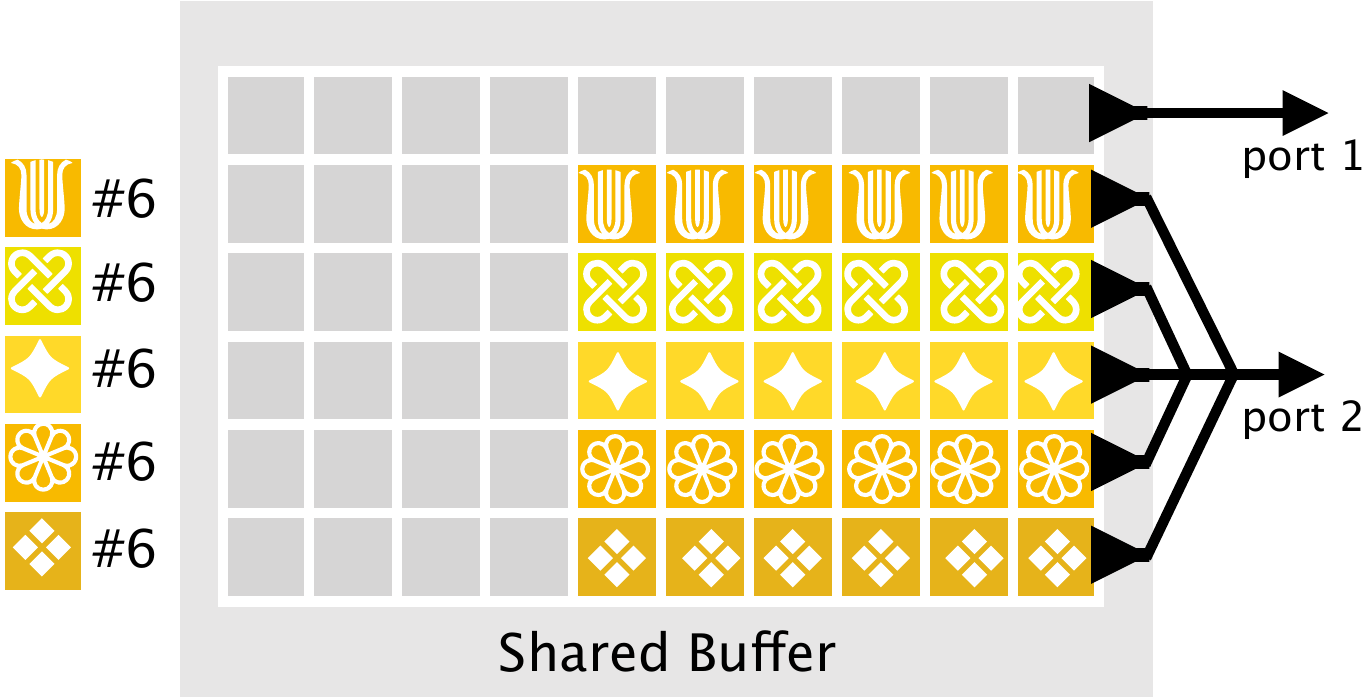}
  \caption[]{\name allocates less space per queue if those share the service rate of a single port.}%
  {{\small }}
  \label{fig:example6}
 \end{subfigure}
  \hfill
  \begin{subfigure}[t]{0.32\textwidth}
  \centering
  \includegraphics[width=\textwidth]{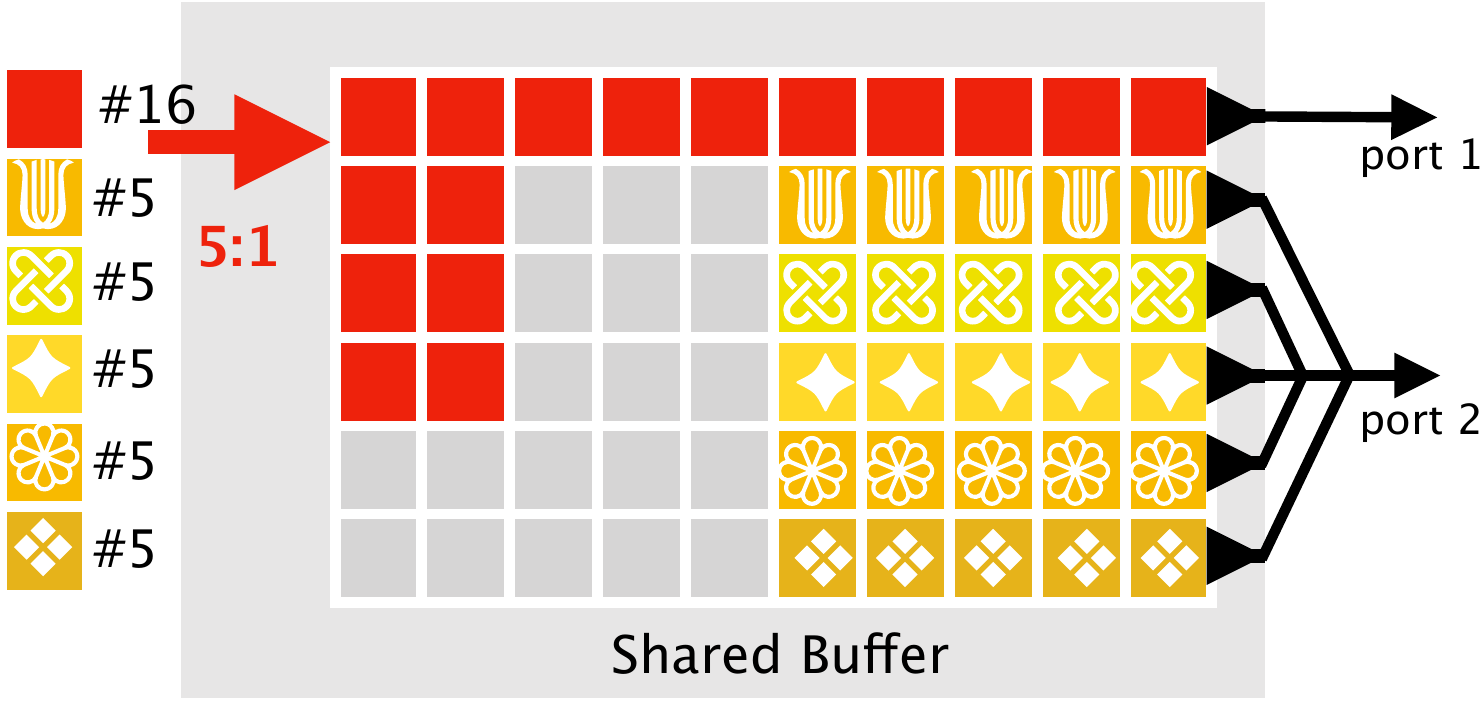}
  \caption[]{The burst is absorbed despite the low aggregate dequeue rate.  }%
  {{\small}}
  \label{fig:example7}
 \end{subfigure}
\caption[]{\name provides minimum buffer guarantees in steady state (a) and burst-tolerance guarantees in transient state (b,c).}
\label{fig:prob2}
\end{figure*}

To sum up, DT has unpredictable transient state.
In effect, the time frame during which a queue experiences drops \emph{before} it allocates its fair share of the buffer (steady-state allocation) can be arbitrary long. Thus, DT cannot guarantee the absorption of a burst.


\section{Overview}
\label{sec:overview}

As we showed in \S\ref{ssec:dt3}, DT fails to offer predictable buffer allocations and thus any performance guarantee.
In this section, we describe \name: a novel buffer management scheme which -unlike DT- manages to control both steady- and transient-state allocation.
We explain how \name addresses DT's limitations by
running it on the same example scenarios we used in \S\ref{ssec:dt3}.

\myitem{\name dynamically bounds the buffer allocation in steady-state.}
\name prevents traffic of any priority from monopolizing the buffer by dynamically bounding buffer usage per priority.
In effect, \name guarantees a minimum allocation to both priorities.
\name is not equivalent to statically allocating space to a single queue (complete partitioning) or to a group of queues (application pools) as it does not statically reserve buffer.

To illustrate the difference between the steady-state allocations of \name and DT we use \fref{fig:example3}, which shows \name's allocation under the same scenario we used for DT in \fref{fig:period3}.
Unlike DT, which decreases the buffer space occupied by high-priority traffic proportionately to the number of low-priority queues, \name bounds low-priority (yellow) queues to $15$ packets on aggregate, equally shared across the three queues.
As a result, the high-priority (red) queue can use $30$ packets of buffer. In \S\ref{sec:design}, we show that \name makes the buffer that high-priority occupies independent of the number of non-empty low-priority queues.
Note that no allocation is static: if the high-priority queue does not use/need its maximum buffer occupancy, the yellow low-priority will get more buffer.

\myitem{\name makes bursts first-class citizens in the buffer by minimizing transient-state drops.}
\name is able to offer burst-tolerance guarantees by allocating buffer such that there is always a combination of \emph{(i)} enough unoccupied buffer space; and \emph{(ii)} adequately high aggregate dequeue rate (\ie the buffer can be emptied fast enough) to accommodate a given burst. Indeed, both these factors are critical for a burst to be absorbed.
On the one hand, an incoming burst can be absorbed independently of the free buffer space at its arrival, iff the aggregate dequeue rate of the allocated buffer space is at least as high as the enqueue rate~\footnote{We are, of course, referring to a burst with size smaller than the total buffer.}. On the other hand, an incoming burst can also be absorbed independently of the aggregated dequeue rate of the buffer at its arrival, iff the unoccupied buffer is sufficient to directly accommodate it.
\name achieves a balance between the two extremes as we explain in \S\ref{sec:design}, which allows it to achieve
high throughput while proving burst-tolerance guarantees.

To illustrate the difference in \name's allocation to that of DT we use \fref{fig:example6} and \fref{fig:example7} which show \name's allocation before and after the arrival of a burst. We again consider the same example as in \fref{fig:example4} and \fref{fig:example5} for DT.

In \fref{fig:example6}, \name detects that the aggregate dequeue rate is inevitably low (queues share a single port) and limits each low-priority queue to 6 packets, effectively leaving an incoming burst enough free buffer to be stored.
As a result, when the 5:1 high-priority burst arrives (\fref{fig:example7}), the buffer can reach steady state fast enough to avoid transient drops.

\vspace{-0.5mm}

\section{Design}\label{sec:design}
\vspace{-0.5mm}

Having explained \name's high-level properties (\S\ref{sec:overview}), we now describe \name in detail.
We elaborate on \name's threshold calculation (\S\ref{ssec:pc}), before we explain its consequences in \name's  performance and guarantees (\S\ref{ssec:guarentees}). 
Finally, we explain how \name's design applies to a single-queue-per-port scenario.

\vspace{-2mm}
\subsection{\name's workings}\label{ssec:pc}
\name limits the buffer space each queue can use depending on both queue-level and buffer-level information.
Particularly, as shown in Eq.~\ref{plast}, \name's per-queue threshold equals the product of: \first an $\alpha$ value assigned to the class that the queue belongs to: $\alpha_c$; \second the inverse number of congested 
(non-empty) queues of the priority (low or high) that the class belongs to: $\frac{1}{N_p(t)}$; \third the per-port-normalized dequeue rate of this queue :$\gamma_{c}^{i}(t)$; and \fourth the remaining buffer space: $B-Q(t)$.

\begin{equation}
T_c^i(t) = \alpha_c  \cdot \frac{1}{N_p(t)}  \cdot \gamma_{c}^{i} (t) \cdot (B-B_{oc}(t))
\label{plast}
\end{equation}
\vspace{-1mm}

\begin{equation*}
\begin{aligned}
 &N_p(t): \text{Number of congested queues of priority p at time $t$} \\
\vspace{-1mm} &B-B_{oc}(t): \text{remaining buffer} \\
\vspace{-1mm} &\text{$\gamma_{c}^{i}(t):$ per-port-normalized dequeue rate of $q_{c}^{i}$ } \\
\end{aligned}
\end{equation*}

\myitem{\name on a single queue per port.}
Eq.\ref{plast} can be naturally adapted to work in a deployment where only a single queue is available per port.
Particularly, $\gamma_{c}^{i}$ will always be $1$ and $N_p(t)$ will correspond to \emph{all} congested queues in the buffer.
Thus, the threshold of a packet of class $c$ destined to port $i$ will depend on the $\alpha_c$, the \emph{total} number of congested queues and the remaining buffer space.
In essence, \name applies \emph{different thresholds} for packets that are mapped to the same queue.

\name's threshold calculation differs from that of DT (Eq.~\ref{eq:dt}) by two factors: \first
 $N_p$; and \second $\gamma_{c}^{i}$.
We explain how each of those differentiates \name's buffer allocation below.

\myitem{$\mathbf{N_p(t)}$ bounds steady-state allocation.}
\name divides per-queue thresholds with ${N_p}$: the number of congested (non-empty) queues of the priority that the class belongs to, as seen in Eq.\ref{plast}. 
The consequence of this factor to \name's allocation is twofold: \emph{(i)} it bounds per-class and per-priority occupancy; and \emph{(ii)} it allows weighted fairness across classes of the same priority.

First, dividing by ${N_p}$ prevents
any single \emph{class}, and any single \emph{priority} from monopolizing the buffer.
As more queues of the same class (or priority) use the buffer, the threshold of each of them decreases, effectively setting an upper bound to the per-class occupancy to $\frac{\alpha_c}{1 + \alpha_c}$ of the total buffer and one to the per-priority occupancy to $\frac{\alpha_p}{1 + \alpha_p}$ of the total buffer, where $\alpha_p$ is the highest alpha of the priority.
As a result, the overall buffer occupancy ($B_{oc}(t)$) is also upper-bounded, as shown in Eq.\ref{eq:BufferOccupancyPCLimit} where $\alpha_L$ and $\alpha_H$ are the maximum $\alpha$ values of the classes of high and low priorities, respectively.
Observe that the maximum aggregate buffer allocation of \name is independent of the number of congested queues. Consequently, the minimum buffer available for a high-priority class is also independent of the number of queues or low-priority classes in the buffer and vise versa.

Other than bounding allocation, dividing by ${N_p}$ offers weighted fairness across classes of the same priority.
Namely, the buffer occupied by a priority is split into classes proportionately to their $\alpha$ values.
As a result, if the operator wishes to favor a traffic class
among those that belong to low priority, she can do so by assigning higher $\alpha$ to this class.

\begin{equation}
\label{eq:BufferOccupancyPCLimit}
B_{oc}(t) \le  \frac{\displaystyle B\cdot (\alpha_L+\alpha_H) }{\displaystyle 1+(\alpha_L+\alpha_H)}
\end{equation}

\myitem{$\mathbf{\gamma_{c}^{i} (t)}$ reduces transient state's duration.}
\name allocates buffer to each queue proportionately to its dequeuing rate ($\gamma$).
The $\gamma$ factor, combined with the upper bounds, minimizes the duration of the transient state.
Indeed, given some amount of buffer per priority, \name splits it into queues proportionately to their service rate, effectively minimizing the time it takes for the buffer to be emptied.
In effect, \name reduces the time needed to transition from one steady-state allocation to another.

\subsection{\name benefits}\label{ssec:guarentees}
Having explained \name's properties, we discuss how
those affect \name in performance metrics.

\myitem{\name improves throughput and reduces queuing delays.}
While \name bounds the amount of buffer used, it maximizes its effectiveness to achieve higher aggregated throughput.
Previous research has shown that TCP throughput benefits from buffer size proportionate to the capacity of the bottleneck link~\cite{ganjali2006update,dhamdhere2006open,redux_ccr_2019}.
Reflecting this observation to the buffer management scheme \name multiplies the thresholds with the queues' dequeue rate, effectively benefiting throughput and reducing buffer pressure while being agnostic to the scheduling algorithm used.
Indeed, \name allocates on average less buffer than DT.
Comparing Eq.\ref{plast} with Eq.\ref{eq:dt} we observe that the added factors decrease the allocated buffer.
As a result, \name keeps queuing delays lower than DT and
the buffer ready to absorb bursts, while not sacrificing throughput.

\myitem{\name guarantees the absorption of a given burst.}
A burst is characterized by its incoming rate $r$ (normalized) and duration $t$~\cite{choudhury1998dynamic}.
Whether a burst will be absorbed depends on: \first its incoming rate ($r$); \second the state of the buffer at its arrival (steady state); and
\third the buffer's ability to dequeue fast (transient state).
We can configure \name to provide two types of guarantees.

First, \name can guarantee that a burst of a given incoming rate ($r$) will be absorbed without any transient losses.
Concretely, the incoming traffic will provably occupy the fair steady-state buffer space that corresponds to its queue before experiencing any drops.
Indeed, Eq.~\ref{rate} shows the maximum $\alpha$ with which the low-priority classes would need to be configured to allow a burst with an incoming rate of $r$ to pass.

\begin{equation}
\label{rate}
\alpha_L \le \frac{1}{r-2}
\end{equation}

Second, owing to its strategic steady-state and transient-state allocation, \name can guarantee that a burst of a given incoming rate $r$ and duration $t$ will be \emph{fully} absorbed\footnote{Of course we refer to bursts that are smaller than the buffer.}.
Concretely, \name guarantees that such a burst will provably experience no drops if $\alpha_H$ and  $\alpha_L$ \ie the maximum $\alpha$ values of high and low priorities respectively are set according to Eq.~\ref{alphaHigh} and ~\ref{alphaLow}.
Observe that neither is dependent on the number of queues occupying the buffer or
their class/priority.

\begin{equation}
\label{alphaLow}
\alpha_L \le \frac{B}{(r-2)\cdot t} -1
\end{equation}

\begin{equation}
\label{alphaHigh}
\alpha_H > \frac{1}{\frac{B}{(r-1)\cdot t \cdot (1+\alpha_L)}-\frac{r-2}{r-1}}
\end{equation}

\myitem{Sketch of proof.}
While we moved the full proof to Appendix~\ref{sec:guarantees}, we include the key intuition below. 
The proof is centered around the time at which a hypothetical burst $(r,t)$ will begin to experience drops, say $t_1$.
$t_1$ plays such a critical role in burst absorption as $t_1$ must be greater than the duration $t$ of a burst for the latter to be absorbed. We first express $t_1$ as a function of $\alpha_L$, $\alpha_H$ and the arrival rate $r$ and derive an upper bound on $\alpha_L$. Indeed, the use of $N_p$ and $\gamma_c^i$ in \name's allocation eliminates the dependency on the state of the buffer. As a result, we are able to obtain guarantees.

Observe that traditional buffer management schemes can only provide such guarantees by statically reserving buffer space for a given burst. Even DT, which is assumed to be more dynamic, would have to limit \emph{all} other priorities extremely, effectively assuming the worst-case scenario all the time to provide burst-tolerance guarantees.

\section{Hardware Design} \label{sec:hwdesign}

Having explained the benefits of \name, 
in this section, we demonstrate its deployability.
To this end, we first explain how we implemented \name on a protocol-independent switch (PISA) (\S~\ref{ssec:pisa}). 
Next, we describe how we can approximate \name's behavior on any device supporting DT, by dynamically adapting the $\alpha$ (\S~\ref{ssec:pca}).

\subsection{\name on PISA}\label{ssec:pisa}

\begin{figure}
 \centering
  \includegraphics[width=\linewidth]{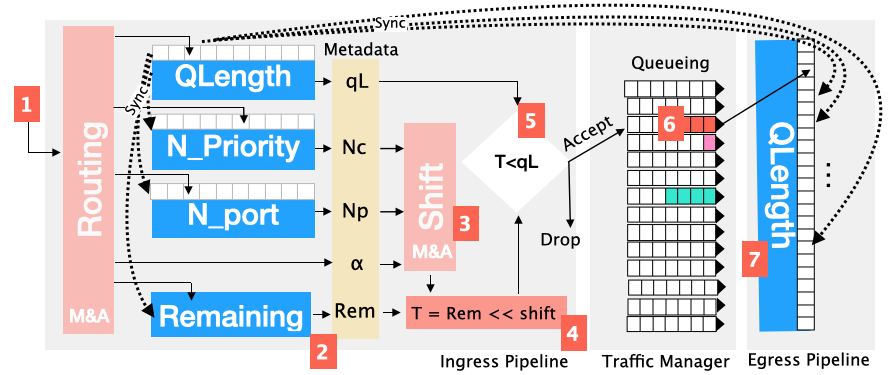}
\caption[]{ \name's hardware design: \name's logic is fully implemented in the programmable ingress and egress pipelines, because the actual buffer management mechanism is a fixed API. Queue lengths are synced from the egress to the ingress pipeline using specially-crafted SYNC packets. }
\label{fig:hw}
\vspace{6pt}
\end{figure}

Naturally, one would implement \name on the Traffic Manager (TM), which is responsible for managing the buffer. Yet, this is not possible as the TM is not programmable~\cite{sharma2020programmable}. Thus, we implemented \name exclusively on the ingress and egress pipelines, which creates three challenges: 
\begin{enumerate}[leftmargin=18pt,topsep=0pt,itemsep=-2pt]
  \item[\restr{1}] Deciding whether a packet should be buffered requires comparing the corresponding queue length with a threshold.
Yet, queue lengths are only available in the egress pipeline; thus, after a packet has been buffered~\cite{sharma2020programmable}.
  \item[\restr{2}] Calculating \name's thresholds requires aggregated metrics over multiple queue lengths, \eg remaining buffer. Yet, accessing multiple values of a single memory block is not possible in PISA switches~\cite{ben2018efficient,chen2018catching}.
  \item[\restr{3}] Calculating \name's thresholds requires floating-point operations, which is not supported by PISA.
\end{enumerate}
Next, we describe \name's high-level hardware design and a packet's journey before we describe how it addresses each of the aforementioned challenges.

\myitem{\name's high-level design}
We built \name upon five main components: four register arrays and two Match \& Action (M\&A) table as shown in Fig.~\ref{fig:hw}
We use the register arrays to keep the required state for deciding whether a packet should be buffered or dropped.
This decision needs to be taken in the ingress pipeline, \ie before the TM accesses the packets.
The aforementioned state includes
the instantaneous remaining buffer (\emph{Remaining}) the number of congested queues per port and priority (\emph{N\_Port}, \emph{N\_Priority}) as well as the queues' length (\emph{Qlength}).
We use a M\&A table (\emph{Routing}) to map a packet based on its destination and priority tag to a port and queue for transmission as well as to multiple \name-specific fields.
Finally, we use another M\&A table (\emph{Shift}) to approximate the required floating-point operations.

\myitem{Packet's journey} Upon arrival \circled{1}, a packet's destination and priority tag are matched against the \emph{Routing} table to multiple action parameters: an $\alpha$ value and three indexes. 
These indexes are used to read the relevant information about the state of the buffer (\eg corresponding queue length from the \emph{Qlength} array or the number of congested queues of a specific priority from the  \emph{N\_Priority}, etc.) \circled{2}.
This information is used to find the required number of shifts \circled{3} to apply to the remaining to calculate the threshold of the corresponding queue \circled{4}.
If the threshold is higher than the corresponding queue's length \circled{5}, the packet is enqueued \circled{6}.
While being at the queue, the packet writes the queue length of its queue to the \emph{Qlength} array in the egress \circled{7}.

\myitem{Queue lengths available to the ingress pipeline}
The length of any given queue is only available as a metadata field to packets that have been enqueued, thus while they traverse the egress pipeline.
Yet, \name requires visibility of queue lengths in the ingress.
To address this, we transfer queue lengths from the egress to the ingress in two steps.
First, we create a register array, which resides in the egress pipeline and stores the length of every queue in the device.
Each index in the array corresponds to a queue, namely a pair of port and traffic class.
Each packet traversing the egress pipeline triggers an update on the value corresponding to the length of the queue it belongs to.
Second, we maintain a copy of this register array in the ingress to make it available to \name's logic. To keep the copy up-to-date, we asynchronously generate specially crafted packets, namely \emph{SYNC} packets. These packets read the queue lengths from the egress register, re-enter in the ingress pipeline via recirculation, and copy the read values to the ingress register array, as shown in Fig.\ref{fig:hw}.
Due to the PISA constraints, which prevent accessing multiple values of a register array in a single pipeline pass, copying all values in one pipeline pass is not possible. Instead, each \emph{SYNC} packet recirculates as many times as queues there are in a device.
While \emph{SYNC} packets need to be sent frequently to allow real-time visibility over the queue lengths, the overhead is minimal as the traffic generator of the device itself can create them, and they use a special pipe that is reserved for recirculation\cite{sharma2020programmable}.

\myitem{Calculating aggregated metrics}
Other than the length of the queue of interest, calculating \name's thresholds requires visibility over:\first its normalized dequeue rate; \second
the number of congested queues of the same priority; and \third the remaining buffer space, as seen in Eq.~\ref{plast}.
These metrics need to be dynamically calculated based on all queues' instantaneous lengths.
Doing so is challenging for three reasons.
First, the dynamic calculation requires accessing multiple values in the same array \eg the number of active queues per port.
Second, it requires accessing selective indexes of the array, namely those corresponding to the subset of queues of interest \eg number of controlled queues per priority.
Third, the result of this calculation needs to be available in the ingress pipeline.
We addressed these challenges again using \emph{SYNC} packets, which read a subset of the indexes of the egress register array,
recirculate and write the aggregated results in the ingress register arrays.
In particular, we use three types of such packets.
First, \emph{SYNC} packets copy queue lengths from the egress to ingress (as described above) and update the \emph{Remaining} register array as they anyway traverse all indexes. 
Second, \emph{SYNC} packets count the congested queues per port, which is equivalent to the normalized dequeue rate per queue given the scheduling algorithm. Each \emph{SYNC} packet updates a single port's count with the number of queues above a threshold.
Finally, \emph{SYNC} packets count the congested queues per priority.
All \emph{SYNC} packets contain in their custom header the indexes from which they start and finish reading from \emph{Qlengths}, the index at which they write and their pivot (indexes they skip).

\myitem{Approximating floating-point operation}
Even after having all required information available,
\name needs to multiply the remaining buffer with other factors (\ie the reverse of the number of congested queues $\frac{1}{N(t)}$, the reverse of the number of active queues per port $\frac{1}{n(t)}$ and the $\alpha$) to calculate the thresholds.
Yet, PISA does not allow floating-point operations.
To address this issue, we shift the remaining space value so many times as the logarithm of two of the product of all the factors mentioned above.
The calculation of the number of required shifts is not done in the data-plane.
Instead, we pre-calculate it for all possible values and store all the results using match-action rules matching on three values $\alpha$, $Np$, and $Nc$.
Observe that all three values are discrete and bounded, so the number of required rules is manageable.
As an intuition, $n$ is in the range of $2-8$;
there are only a couple of possible $\alpha$ (8 for Tofino),
and a few decades congested queues at most.

\subsection{\name on top of DT (\app)}\label{ssec:pca}
We can approximate \name's behavior (\app) by periodically reconfiguring $\alpha$ per queue according to the buffer occupancy. 
Recall from Eq.~\ref{plast} that \name's formula deviates from DT's $\alpha$, say $\alpha_{dt}$, such that  
$\alpha_{dt}=\alpha_c  \cdot \frac{1}{N_p(t)}  \cdot \gamma_{c}^{i} (t)$. 
Both the number of congested queues $N_p$ and the normalized dequeue rate ($\gamma$) change over time; thus, we need to monitor both variables to calculate $\alpha$. 
Fortunately, chip manufacturers (\eg Broadcom) expose the queue lengths and the remaining buffer (at least as watermarks).
Thus, we can build \app by using a software controller that \emph{(i)} periodically pulls the queue lengths and the remaining buffers; \emph{(ii)} calculates the number of congested queues; and \emph{(iii)} infers the per-queue dequeue rate considering the scheduling algorithm per port and the number of active queues.
The periodicity of the $\alpha$ updates depends on the capabilities of the device in terms of monitoring queues and updating $\alpha$.
In \S\ref{sec:evaluation}, we consider the update period of $1s$ which already brings considerable benefits.
Intuitively, as we increase the frequency of the updates, \app will approach the performance of \name.

Despite its benefits, \app cannot approximate \name if only a single queue is available per port. 
As \app relies on DT it cannot use different thresholds for packets of the same queue. In this case, \app will have the same performance as DT.
Still, provided that traffic is stable, \name's insights can be used to configure static $\alpha$.

\section{Evaluation}
\label{sec:evaluation}

\begin{figure}[h]
\centering
\includegraphics[width=0.9\linewidth]{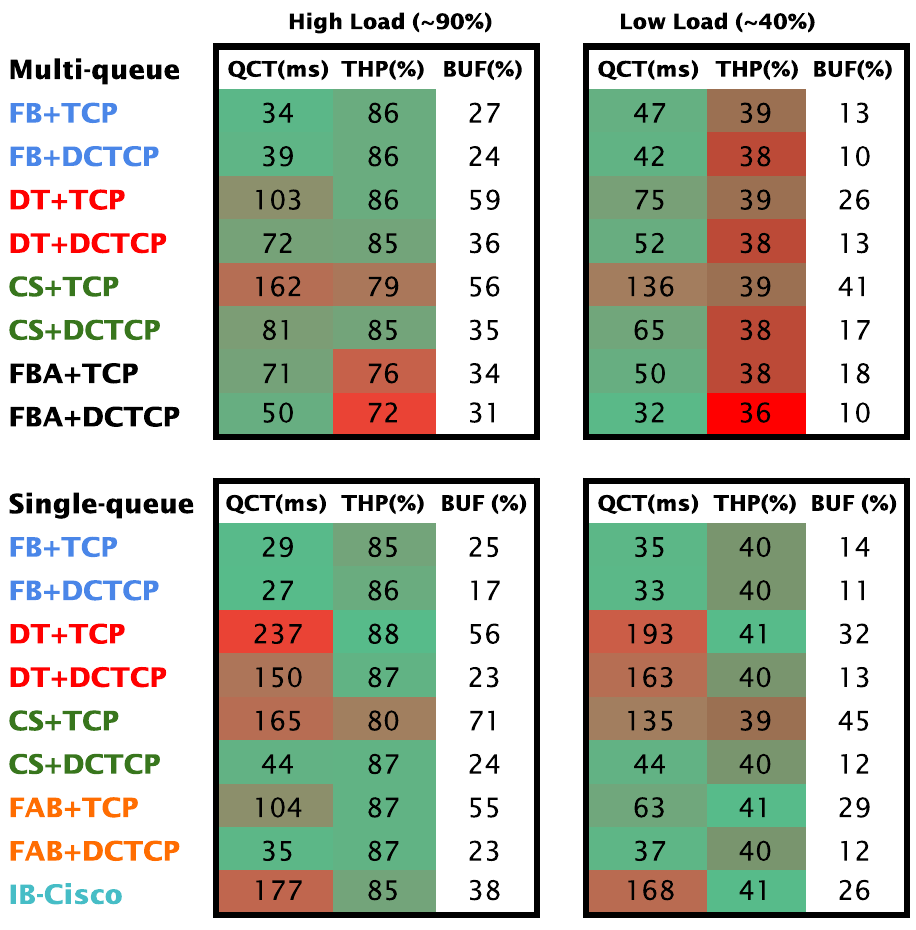}
\caption{Summary of results comparing average Query Completion Time (QCT), throughput (THP), and buffer usage (BUF) across various buffer management schemes (rows) for high and low load (left, right tables) considering multi- and single-queue deployment scenarios (top, bottom tables). Greener cells illustrate better performance. \name (top 2 rows in each table) achieves the best burst absorption measured as QCT without sacrificing throughput. \name's approximation (\app) (bottom 2 rows in the top tables) improves burst absorption but decreases throughput. }
\label{fig:summary}
\end{figure}
\begin{figure*}[t]
\begin{subfigure}[ht]{1\linewidth}
\centering
\includegraphics[trim=0 0 0 0,clip,width=1\linewidth]{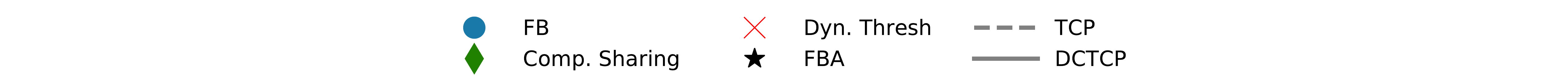}
\end{subfigure}
\begin{subfigure}[ht]{.32\linewidth}
\centering
\includegraphics[trim=0 0 0 0,clip,width=1\linewidth]{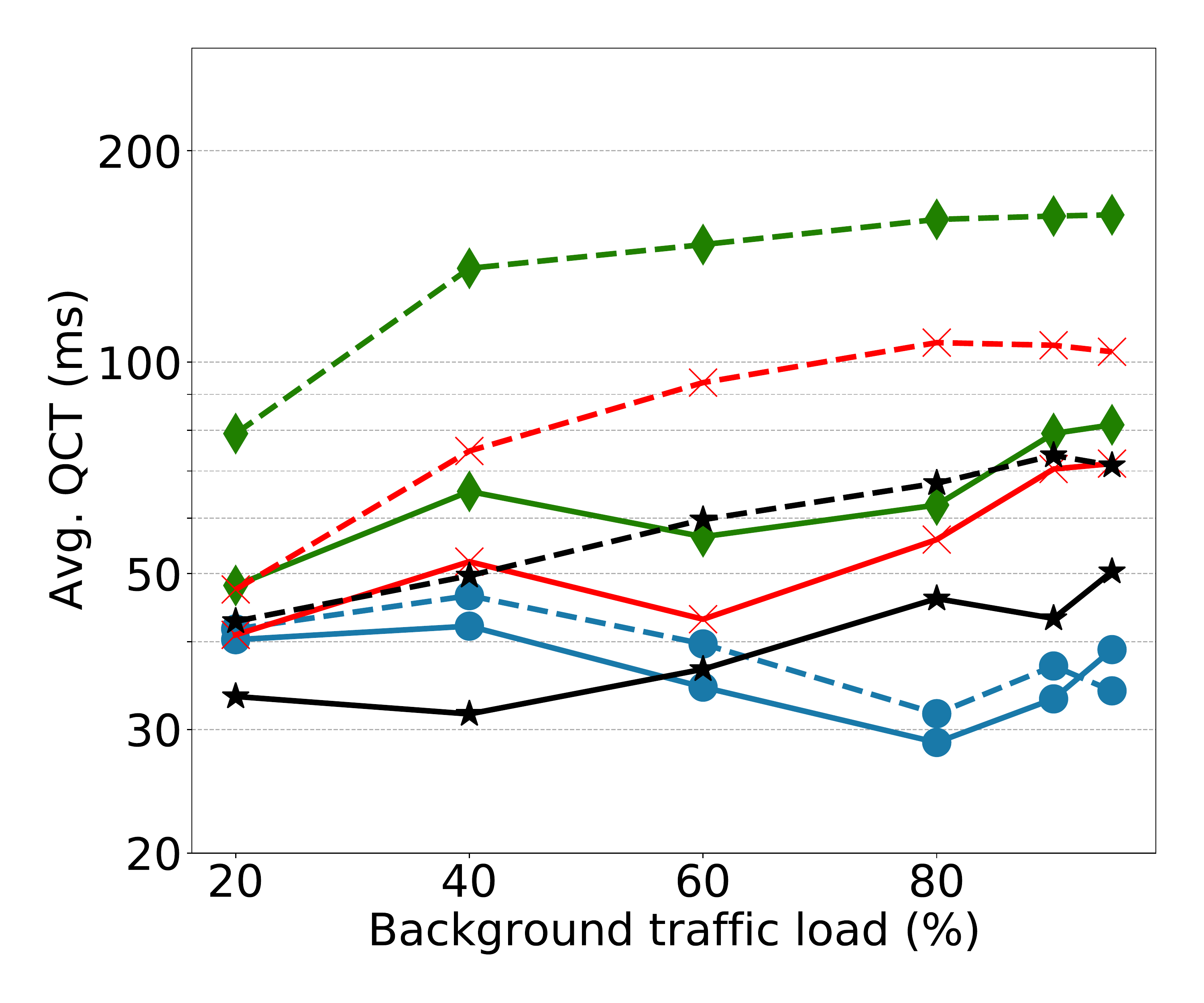}
\caption{}
\label{fig:burst-multi-load}
\vspace{1mm}
\end{subfigure}
\begin{subfigure}[ht]{.32\linewidth}
\centering
\includegraphics[trim=0 0 0 0,clip,width=1\linewidth]{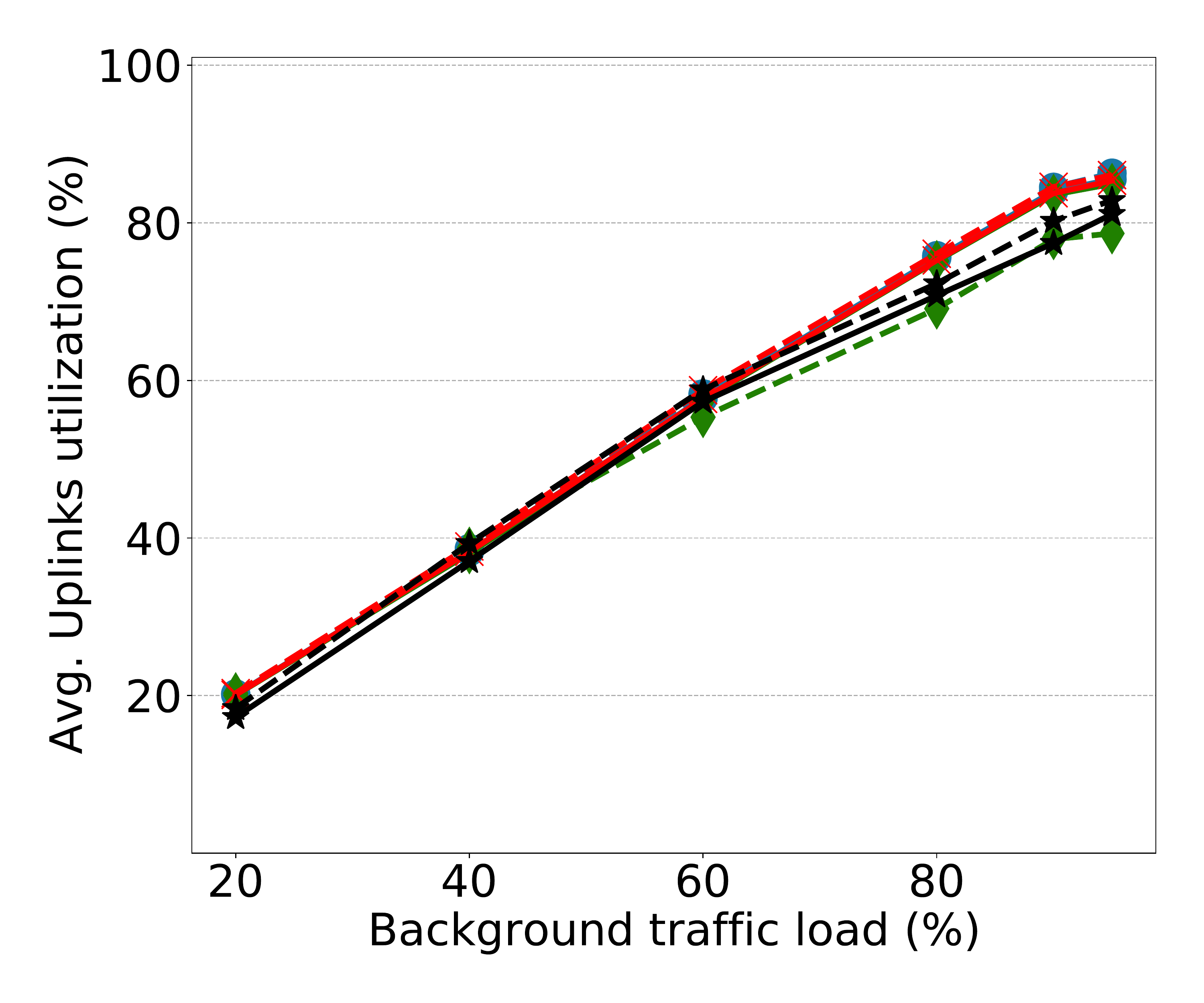}
\caption{}
\label{fig:th-multi-load}
\vspace{1mm}
\end{subfigure}%
\begin{subfigure}[ht]{.32\linewidth}
\centering
\includegraphics[trim=0 0 0 0,clip,width=1\linewidth]{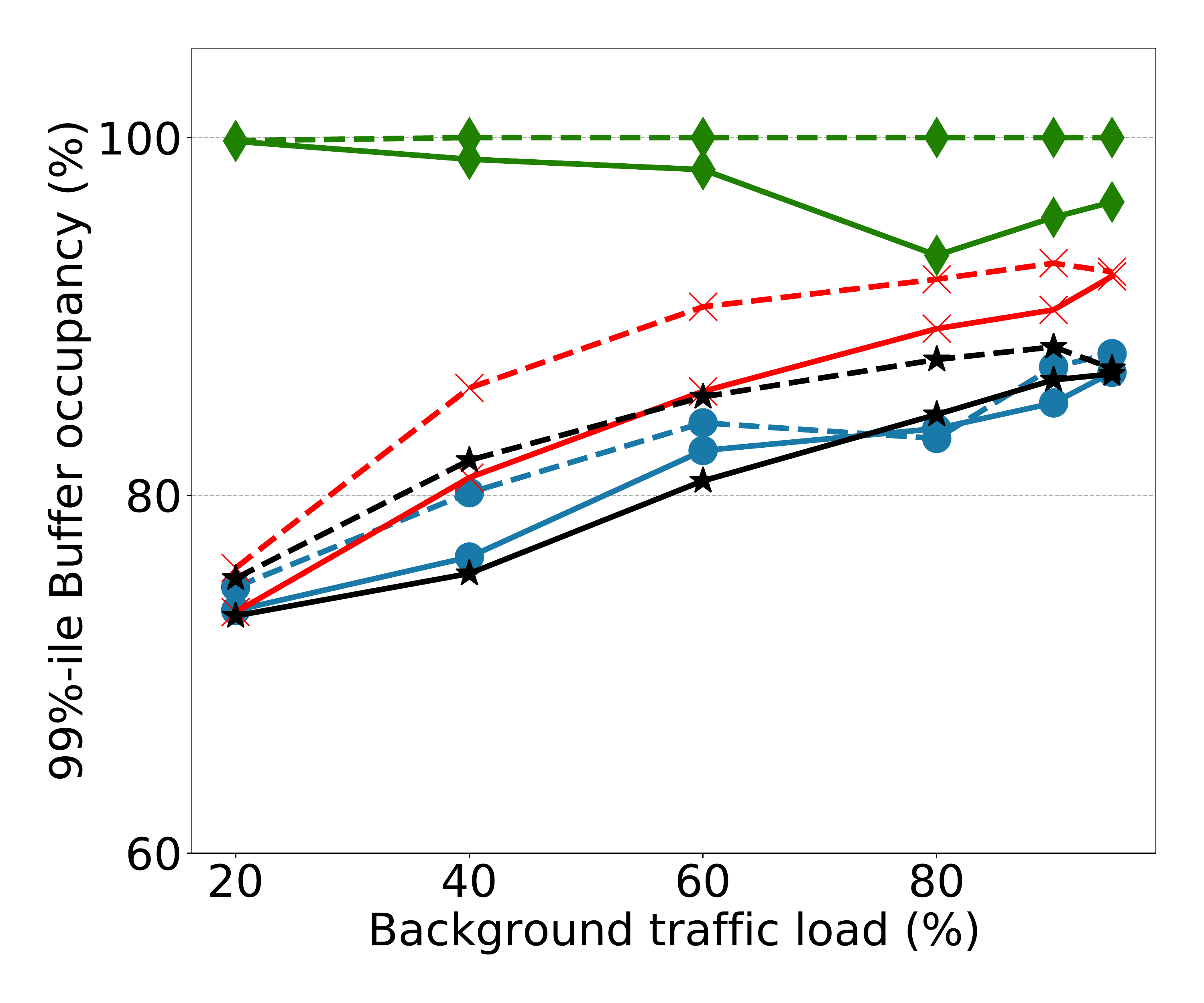}
\caption{}
\label{fig:buffer-multi-99}
\vspace{1mm}
\end{subfigure}
\begin{subfigure}[ht]{.32\linewidth}
\centering
\includegraphics[trim=0 0 0 0,clip,width=1\linewidth]{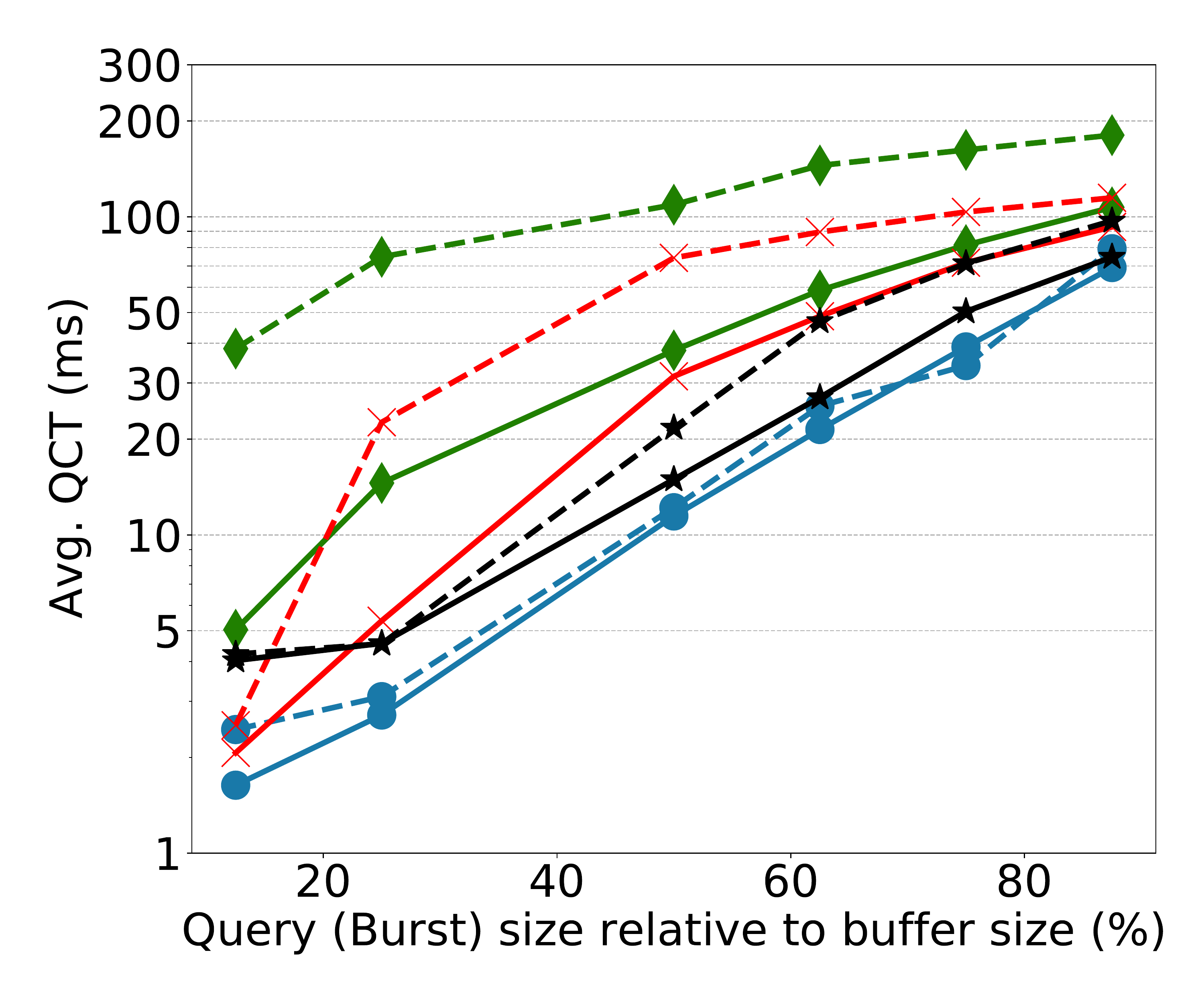}
\caption{}
\label{fig:burst-multi-burst}
\end{subfigure}
\begin{subfigure}[ht]{.32\linewidth}
\centering
\includegraphics[trim=0 0 0 0,clip,width=1\linewidth]{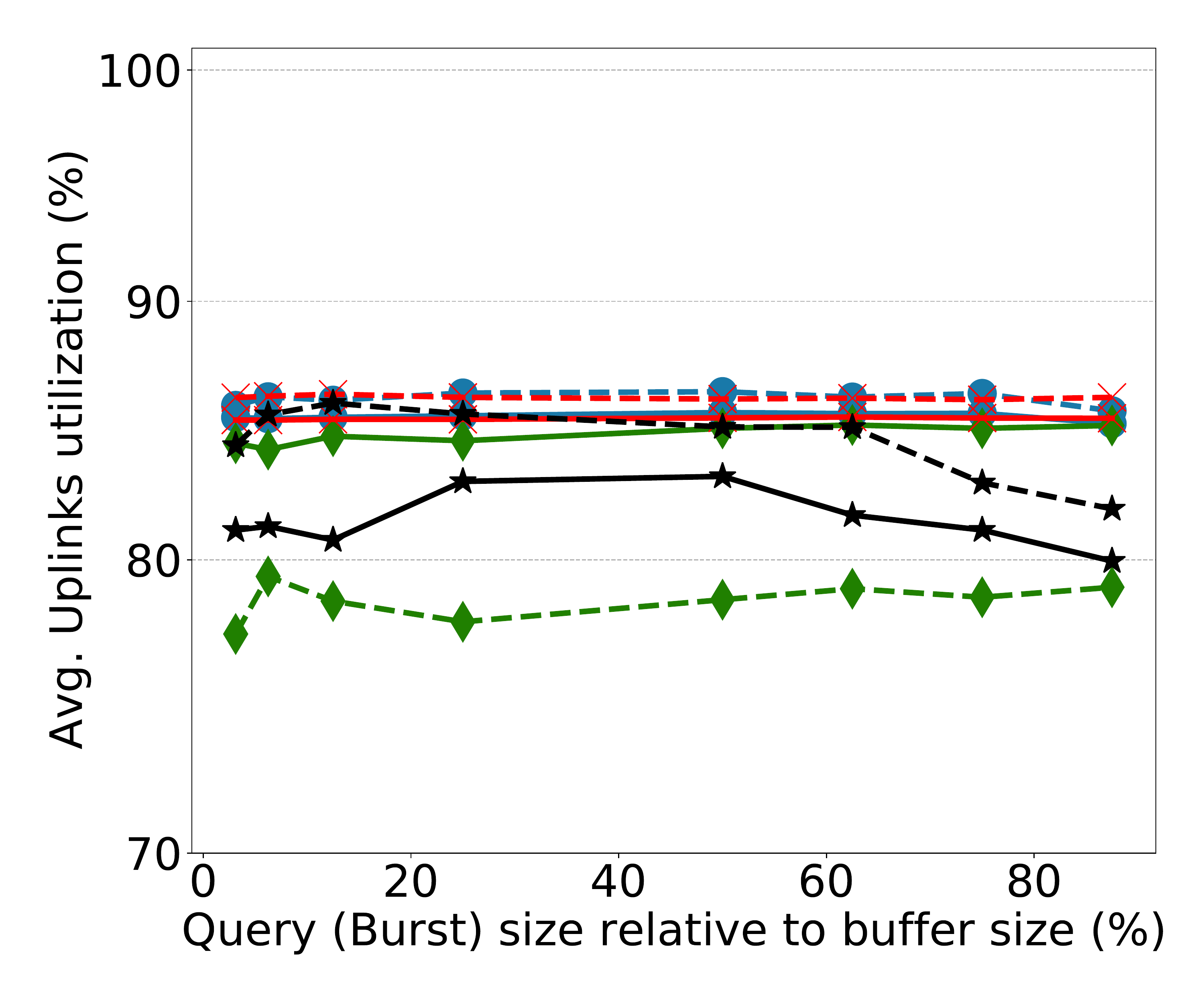}
\caption{}
\label{fig:th-multi-burst}
\end{subfigure}%
\begin{subfigure}[ht]{.32\linewidth}
\centering
\includegraphics[trim=0 0 0 0,clip,width=1\linewidth]{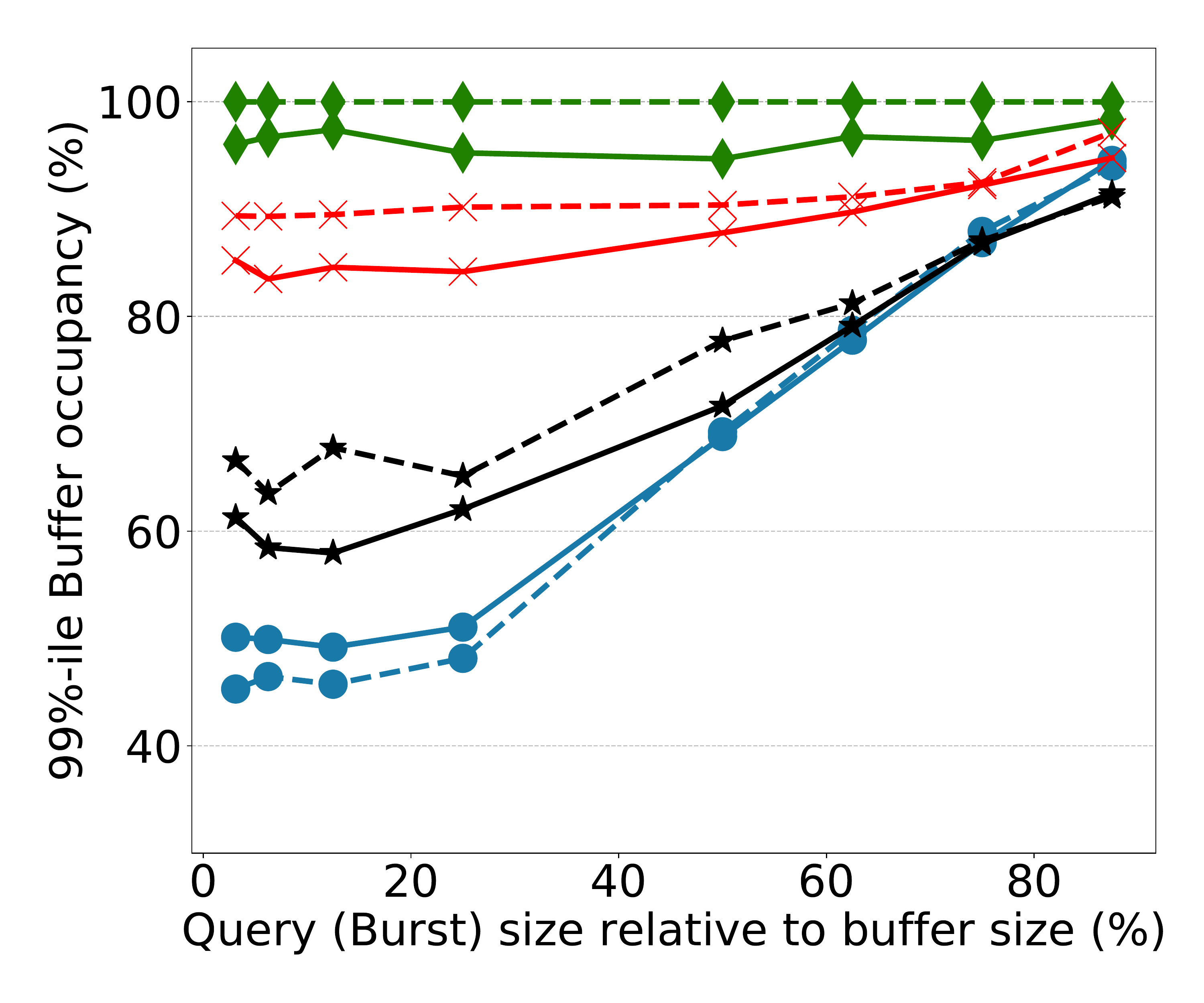}
\caption{}
\label{fig:buffer-multi-99}
\end{subfigure}
\vspace{-1.5mm}
\caption{ \textbf{Multi-queue scenario:} \name outperforms in QCT while achieving on-par throughput compared to \emph{all} other schemes even when they are paired with DCTCP.
\name's 99-th percentile (99p) buffer occupancy increases mostly to accommodate bursts.}
\end{figure*}

We evaluate \name aimining at answering four main questions:
\begin{enumerate}[leftmargin=18pt,topsep=0pt,itemsep=-0pt]
  \item[\q{1}] How does \name compare against other buffer schemes? \label{q1}
  \item[\q{2}] Is \name useful if DCTCP is already deployed? \label{q2}
  \item[\q{3}] Does \name deteriorate performance under low load? \label{q3}
  \item[\q{4}] Is \name useful if traffic is not already marked/classified? \label{q4}
\end{enumerate}

\myitem{}We demonstrate that: (\textit{\textbf{Q1}}) \name outperforms \emph{all} buffer management algorithms in terms of burst tolerance while achieving on-par throughput;
(\textit{\textbf{Q2}}) \name even with TCP improves burst absorption compared to DT with DCTCP by 53\% under high load and by 10\% under low load;
(\textit{\textbf{Q3}}) \name
does not deteriorate throughput even in the absence of bursts or under low load as it does not statically allocate buffer; and (\textit{\textbf{Q4}}) \name
brings even more benefits when the traffic classification is done directly on the device.
We first give a summary of our key findings (\S\ref{ssec:sumary}).
Next, we elaborate on our methodology (\S\ref{ssub:methodology}) and we describe our detailed results (\S\ref{ssub:results}).

\remove{
In our evaluation, we compare \name's performance to alternative buffer management schemes
and TCP versions under various traffic loads. To that end, we focus on \emph{burst absorption} measured as Query Completion Time (QCT), throughput and buffer usage. We assume that traffic is marked and there are multiple queues per port (multi-queue scenario). Next, we remove both assumptions to investigate the usefulness of \name in a single-queue scenario.

}

\subsection{Key Findings}\label{ssec:sumary}
Fig.~\ref{fig:summary} summarizes our key findings in four tables.
Rows correspond to a combination of a buffer management algorithm with a TCP version\footnote{We do not evaluate FAB and IB in the multi-queue scenario as they are not compatible with marked traffic, \ie they classify packets on their own.}. Columns present average metrics for Query Completion Time (QCT), throughput (THP), and buffer utilization (BUF). Greener cells in the first two columns correspond to better performance. All cells in the third columns are white as buffer utilization does not trivially translate to performance.
The two tables at the top of Fig.~\ref{fig:summary} correspond to the
multi-queue scenario: marked traffic and multiple available queues per
port while those at the bottom to the single-queue scenario: single
queue per port. The two tables on the left correspond to the case of relatively high network load ($90\%$) while the right ones to relatively low network load ($40\%$).
\myitem{\name} is the most effective algorithm in burst absorption compared to all alternative buffer management algorithms while achieving on-par throughput.
Notably, this holds even if we compare \name with TCP to alternatives paired with
DCTCP.
\name reduces DT's average QCT by 67\% (69ms)  
under high load (90\%) and by 37\% (28ms) under low load (40\%). 
\name benefits increase further in the single-queue scenario.
In any case, \name achieves on-par throughput, which is (in the worst case) 3\% lower than that of DT.
Interestingly, \name achieves better burst absorption under high load than under low load. This is the case because when the load is lower, \name allocates more buffer per port on average. As a result, the aggregate dequeue rate under low load decreases together with the buffer's ability to fully absorb an incoming burst, as we explain in \S\ref{sec:background}.

\myitem{Dynamic Thresholds} (DT) optimizes throughput by using more buffer.
Indeed, DT with TCP uses twice as much buffer as \name when TCP is used. Still, DT significantly penalizes bursts, achieving the worst performance across \emph{all} alternative buffer management schemes
in the single-queue scenario under high load.
DCTCP combined with the isolation offered by multiple priority queues improve DT's burst absorption capabilities, but it still performs 84\% worse than \name.

\myitem{Complete Sharing} (CS) is equivalent to no buffer management.
Interestingly, CS is effective in absorbing bursts when paired with DCTCP. Indeed, CS with DCTCP reduces QCT by 70\% compared to DT with DCTCP, and it achieves high throughput. Still,
 CS with DCTCP performs 51\% 
 worse than \name in burst absorption. Naturally, CS with TCP under-performs in both burst absorption and throughput.
 In any case though CS allows a single malicious or non-responsive flow to monopolize the buffer, thus cannot be used in practice.

\myitem{\name approximation} (\app) (\S\ref{ssec:pca}) achieves similar performance with \name in burst absorption.
In particular, when DCTCP is used \app increases QCT by 10-11ms compared to \name
but it decreases DT's QCT by 116-118ms.
However, \app sacrifices throughput. Indeed \app reduces \name's throughput by 2-10\%. The reason for this inefficiency is that the $\alpha$ is only re-configured periodically, while the state of the buffer changes much faster. Still as we increase the frequency or as the traffic becomes less dynamic the performance of the approximation will approach that of \name.

\myitem{Flow Aware Buffer} (FAB)~\cite{apostolaki2019fab} outperforms conventional algorithms in QCT, but it does not reach the level of \name.
Indeed, \name reduces FAB's QCT by 72\% with TCP and by 23\% with DCTCP under high load.

\myitem{Cisco Intelligent Buffer} (IB)~\cite{cisco9000} is ineffective in absorbing bursts but achieves high throughput. The main reason for this inefficiency is the extensive use of the buffer. Observe that IB uses almost 2x more buffer on average than \name.
This leaves little space to absorb bursts.

\subsection{Methodology}\label{ssub:methodology}

Having briefly discussed our key insights, we elaborate on the topology, traffic mix, priority assignment and deployment scenarios we used.

\myitem{Topology.}
We evaluate \name's performance in a Leaf Spine topology~\cite{leafspine} of two leaves and two spines with four links connecting each pair of leaf and spine.
We use ECMP to load-balance traffic across uplinks.
We set the buffer size to 1MB and the bandwidth per port to 1Gbps.
Each leaf is connected to 40 servers (oversubscription of 5).
We set $RTT=200\mu s$.
The results we describe are not sensitive to concurrent increases of bandwidth and buffer size.
Naturally, the buffer management is less important, if the buffer size increases
with static aggregate bandwidth.

\myitem{Traffic mix.}
We generate two types of traffic: \first web search (realistic workload based on traffic measurements in deployed datacenters); and \second query traffic (models the user queries of a web application backend).
We selected these two traffic types as examples of non-bursty and bursty traffic, respectively. Our results hold for similar distributions that require both burst absorption and high throughput.

For the former type, we use the flow size distribution from~\cite{alizadeh2011data} and tune the mean of Poisson inter-arrivals such that a $20-90$\% load is achieved.

For the latter type, we assume that a query arrives at each server according to a Poisson process with mean 1 query/second following~\cite{leafspine}. We vary the size of the query from 0 to 90\% of the buffer size.
Burst of zero size corresponds to the case when no high-priority traffic arrives.
Each query consists of a server attached to a leaf requesting a
\emph{Query-Size} file from all the servers connected to the other leaf.
Each request is then responded by 40 servers, each transmitting $\frac{1}{40}$ of the file.
Each pair of servers has a persistent TCP connection as in \cite{leafspine}.
A query is completed when the requester receives the \emph{Query-Size} file.

\begin{figure*}[th]
\begin{subfigure}[ht]{1\linewidth}
\centering
\includegraphics[trim=0 0 0 0,clip,width=1\linewidth]{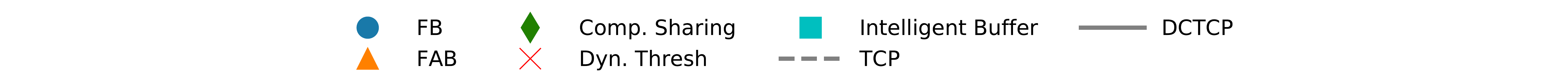}
\end{subfigure}
\begin{subfigure}[ht]{.32\linewidth}
\centering
\includegraphics[trim=0 0 0 0,clip,width=1\linewidth]{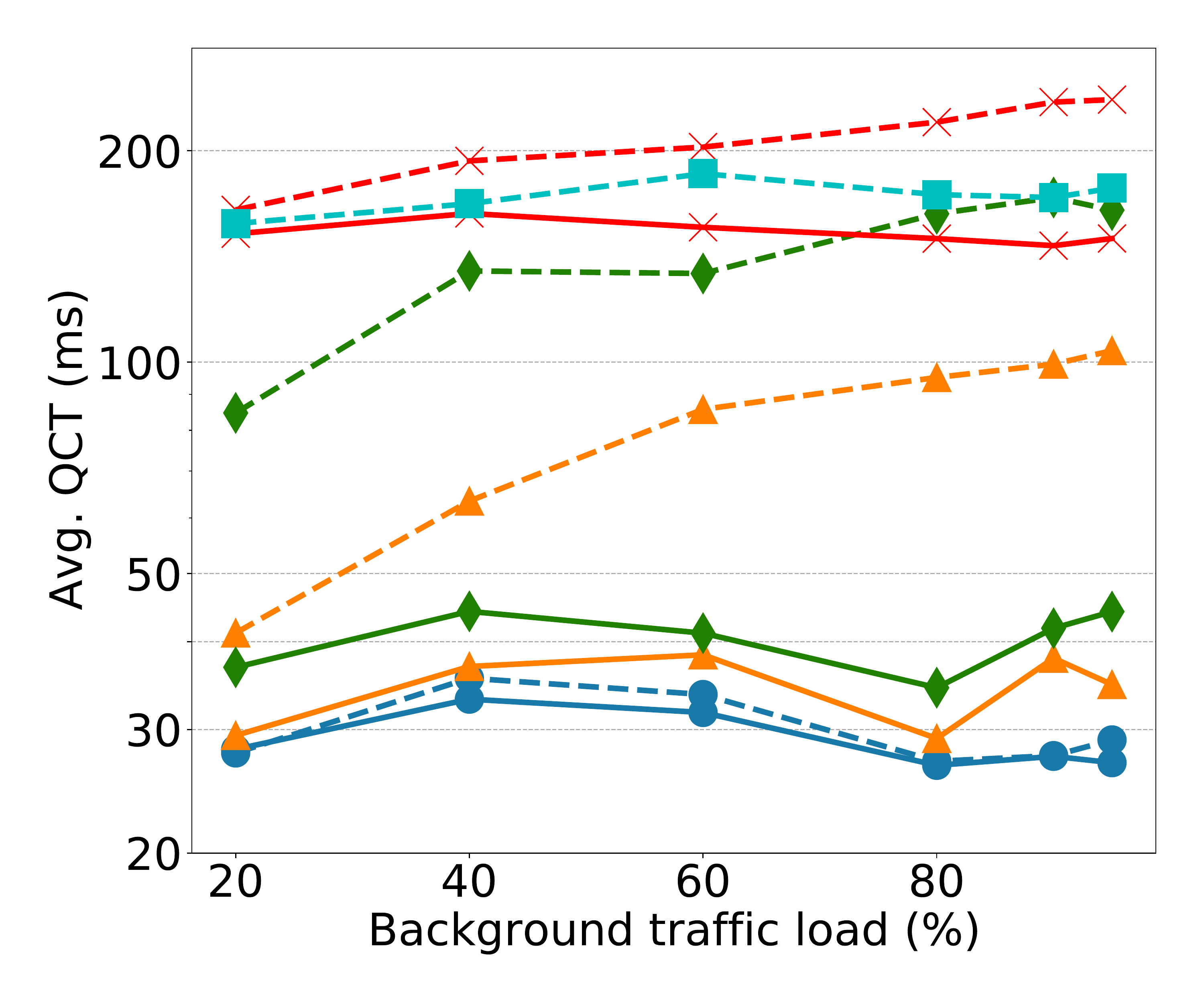}
\caption{}
\label{fig:burst-single-load}
\end{subfigure}
\begin{subfigure}[ht]{.32\linewidth}
\centering
\includegraphics[trim=0 0 0 0,clip,width=1\linewidth]{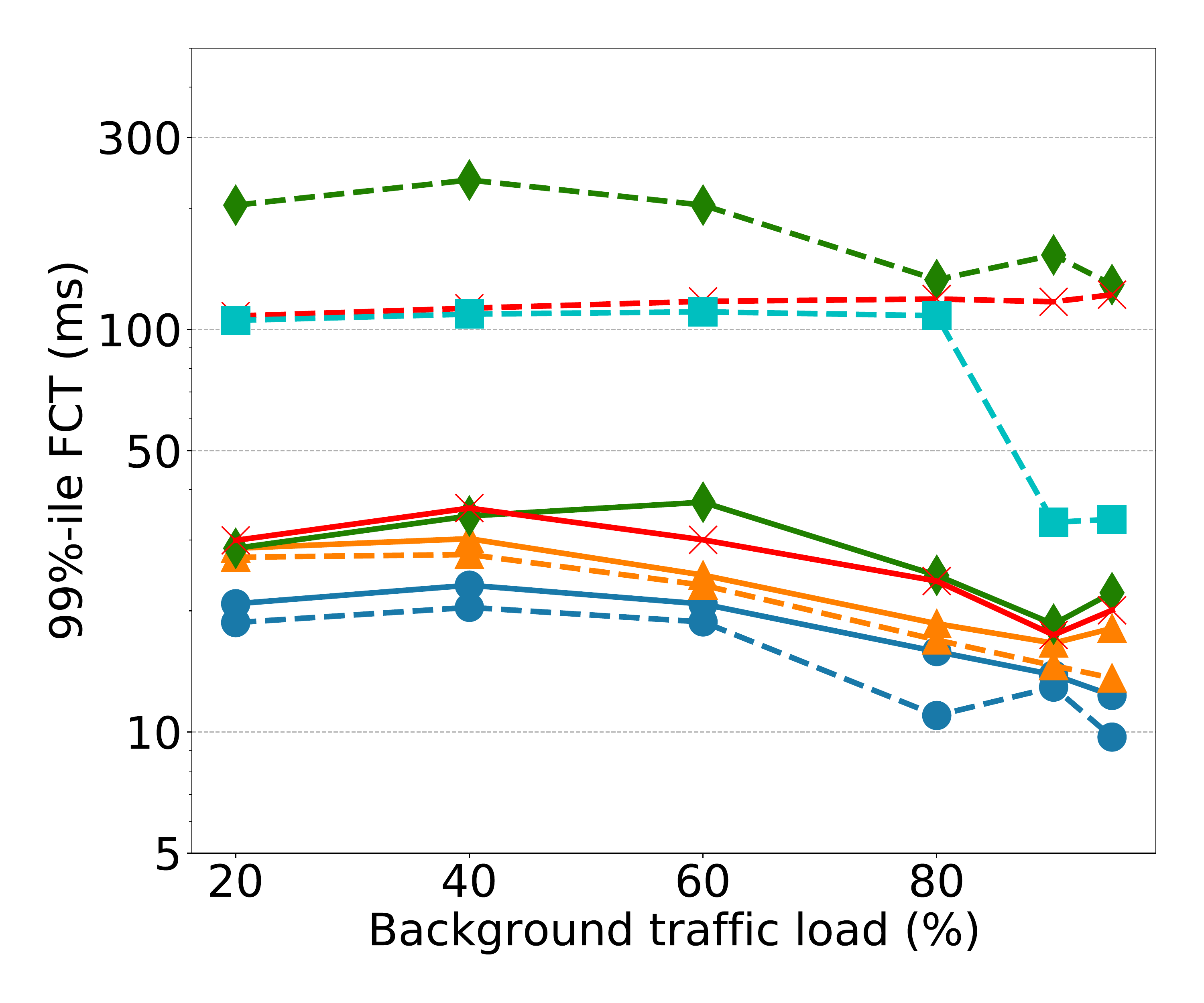}
\caption{}
\label{fig:fct-single-99}
\end{subfigure}
\begin{subfigure}[ht]{.32\linewidth}
\centering
\includegraphics[trim=0 0 0 0,clip,width=1\linewidth]{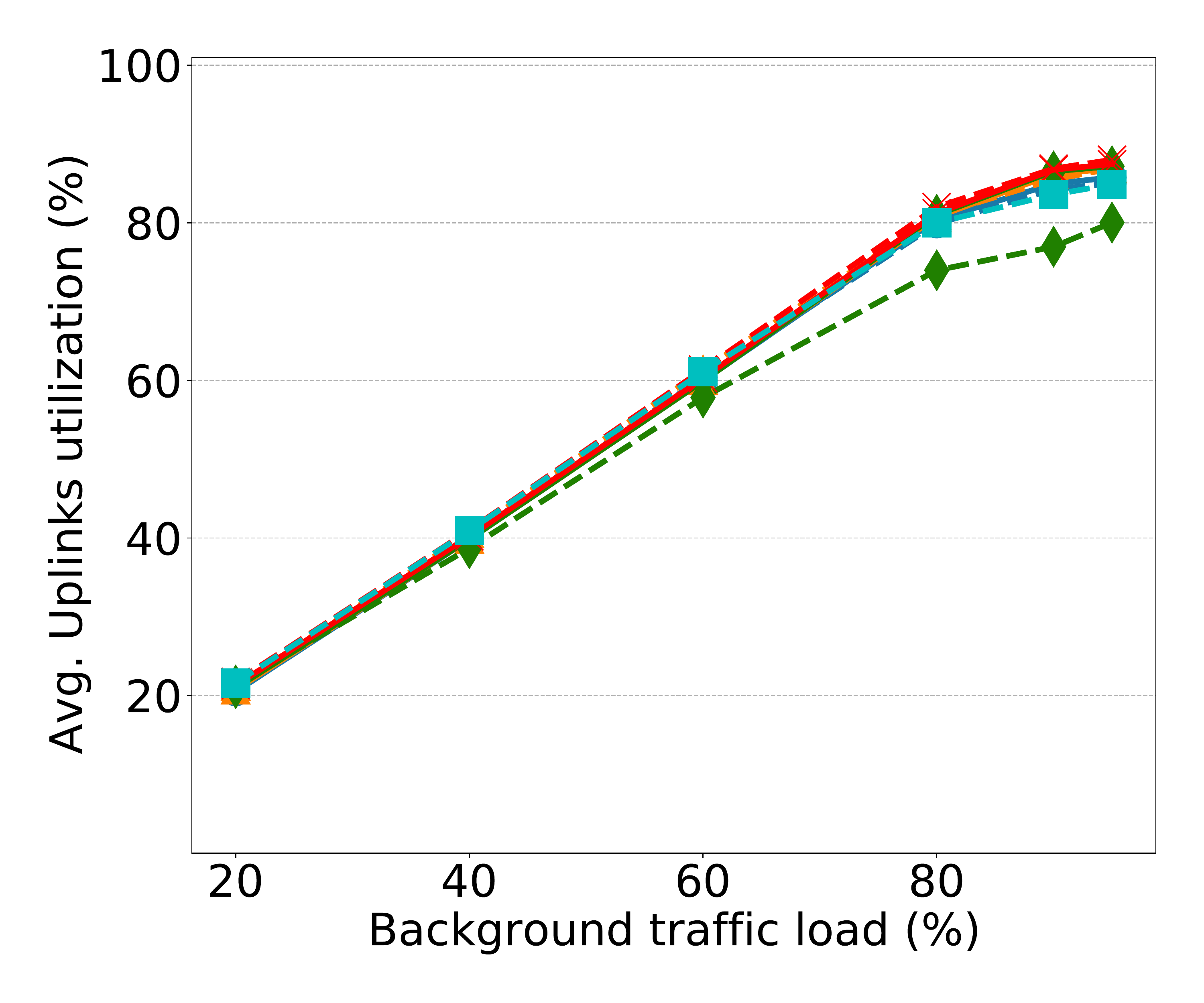}
\caption{}
\label{fig:th-single-load}
\end{subfigure}%
\caption{\textbf{Single-queue scenario:} \name significantly improves QCT and FCT compared to \emph{all} other schemes
under various traffic loads. \name achieves on-par throughput. FAB approaches the performance of \name only when combined with DCTCP. FAB with DCTCP has 14\% higher QCT and 31\% higher FCT compared to \name with TCP, under 60\% load.}
\end{figure*}
\myitem{Deployment Scenarios.}
We consider two deployment scenarios:
\emph{multi-queue} and \emph{single-queue}, which differ on the number of priority queues and the marking scheme.

In the \textbf{multi-queue} scenario, we assume five queues per port and Round-Robin scheduling.
Incoming packets are mapped to a queue according to the tag they carry.
The query traffic is marked with a high-priority tag and the web-search traffic with 4 equally-low-priority tags.
Tags are added by the servers uniformly at random.
In effect, packets belonging to a burst carry a high-priority tag and are mapped to a separated priority queue in the multi-queue deployment scenario.
In this scenario, we compare \name and its approximation with two other buffer management schemes: \emph{(i)} Complete Sharing (CS), which allows queues to grow arbitrarily large in the shared buffer until it is full; and
 \emph{(ii)} Dynamic Thresholds (DT), which we extensively describe in \S\ref{ssec:dt2}. Unlike in the single-queue scenario, for this scenario we do not evaluate
other buffer management algorithms \ie the Cisco Intelligent Buffer (IB)~\cite{cisco9000} and FAB~\cite{apostolaki2019fab} as they use a classifier that overwrites the pre-marked traffic.

In the \textbf{single-queue} deployment scenario, we assume a single queue per port.
Traffic is not marked with priority tags by end hosts,
but a flow classifier is available to \emph{all} buffer management algorithms.
The availability of the classifier allows us to compare more buffer management schemes.
Thus, other than \name, DT and CS we also compare against the Cisco Intelligent Buffer (IB)~\cite{cisco9000} and FAB~\cite{apostolaki2019fab} which require a classifier and cannot use the pre-marked traffic.
In particular, the Cisco Intelligent Buffer (IB)~\cite{cisco9000} combines DT, a fair dropping mechanism, and a priority queue,
while FAB is a newer scheme which prioritizes short flows~\cite{apostolaki2019fab}.

\myitem{Configuration.}
We configure \name with $\alpha_L=0.5$ and $\alpha_H=20$ for short and long flows, respectively in the single-queue scenario; and for low-priority and high-priority traffic, respectively in multi-queue scenario.\footnote{We derive the parameters for \name based on our analysis in Appendix.~\ref{sec:guarantees} optimized for burst sizes $\le75\%$ of the buffer size.}
We configure DT with $\alpha=0.5$ in the single-queue scenario;\footnote{Setting $\alpha=20$ would approximate absence of buffer management.} and with $\alpha_L=0.5$ and $\alpha_H=20$ for low-priority and high-priority traffic, respectively in the multi-queue scenario.
We configure IB with $\alpha=0.5$. IB uses headroom for short flows. IB also uses a separated priority queue for short flows in the single-queue scenario. Finally, IB uses Approximate Fair Dropping (thus cannot use DCTCP).
When DCTCP is used, all queues are RED~\cite{red} with min and max thresholds set to $20$ ($K=20$) following the recommendations in~\cite{alizadeh2011data}.
When TCP is used, all queues are DropTail, except for IB.
TCP \emph{minRTO} is set to $100ms$.

We use the ns-3 Simulator, version $3.31$.\footnote{Our implementation of the evaluated buffer-management algorithms in ns-3 will be made available online.}
We perform $10$ experiments and report average values.

\subsection{Results}\label{ssub:results}

\myitem{\name burst absorption capabilities are independent of the background load.}
Fig.~\ref{fig:burst-multi-load} illustrates the average QCT as a function of the low-priority load, while the burst size is fixed to 75\% of the buffer.
While QCT increases as the load increases for most of the buffer management algorithms,
QCT for \name is consistently low.
\name guarantees high burst absorption even under high load by bounding the buffer usage of low-priority traffic.
DCTCP demonstrates similar behavior. Particularly, when paired with DCTCP both CS and DT have better burst absorption capabilities (lower QCT).
Still, \name demonstrates lower QCT starting from 40\% load compared to \emph{all} alternatives, even when paired with TCP.

\myitem{\name uses more buffer as the burst size increases.}
Fig.~\ref{fig:burst-multi-burst} illustrates average QCT as a function of burst size, while the load is fixed to 90\%.
The benefits of \name in terms of burst absorption increase as the burst size increases but it starts being superior even when the burst size is 10\% of the buffer.
The key reason for this behavior is that \name increases its buffer usage only in case of a burst of high-priority traffic. Indeed, as we observe in Fig.~\ref{fig:buffer-multi-99}, \name's 99p (99-th percentile) buffer occupancy increases as the burst size increases.

\myitem{\name achieves on-par throughput under any background load and buffer size.}
Fig.~\ref{fig:th-multi-load} illustrates the achieved throughput as a function of load on the device, while  Fig.~\ref{fig:th-multi-burst} as function of burst size.
Naturally, the throughput increases with the load for all algorithms as well as for \name.
Observe though that when CS is paired with TCP (not DCTCP) the throughput decreases, because
the buffer allocation of CS is first-come-first-served, thus highly sensitive to bursts and flows' relative order.
Importantly, the approximation of \name (\app)  demonstrates slightly lower throughput when paired with TCP.
This is due to the short duration of burst combined with the infrequent change of $\alpha$, that causes
\app to reduce the buffer given to low-priority for longer than necessary to absorb the burst.

\myitem{In the single-queue scenario \name has higher benefits.}
Fig.~\ref{fig:burst-single-load}, \ref{fig:fct-single-99} illustrate average QCT and 99p FCT as a function of background load, respectively.
\name improves average QCT by 83\% and 99p FCT by 38\%  compared to DT with DCTCP when background load is only 20\%.
All algorithms (other than CS paired with TCP) achieve on-par throughput as we observe in Fig.~\ref{fig:th-single-load}. When paired with DCTCP, FAB and CS also improve DT's performance in terms of QCT and FCT.
Still \name decreases QCT by 12\% and FCT by 24\% under 60\% even when FAB uses DCTCP and \name uses TCP. Of course, \name's performance further improves with DCTCP.

\myitem{\name blindly adheres to the marking, thus erroneous marking might lead to undesirable performance.}
As an intuition, short flows of low-priority traffic might experience high tail FCT if they compete with long flows of the same class. Thus, if the operator wants to avoid it, she should mark all short flows as high priority.
If the operator does not wish to mark packets then she can leave this operation to an classifier residing in the network device as in the single-queue scenario.
Similarly, if the operator configures UDP traffic as high-priority
then that might significantly affect the performance of other high priority-traffic.

\begin{figure}
 \centering
  \begin{subfigure}[b]{0.57\linewidth}
  \centering
  \includegraphics[width=\linewidth]{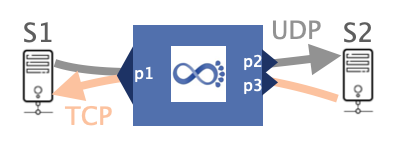}
  \caption[]{}%
  { }
  \label{fig:testbed}
 \end{subfigure}
  \begin{subfigure}[b]{0.41\linewidth}
  \centering
  \includegraphics[width=\linewidth]{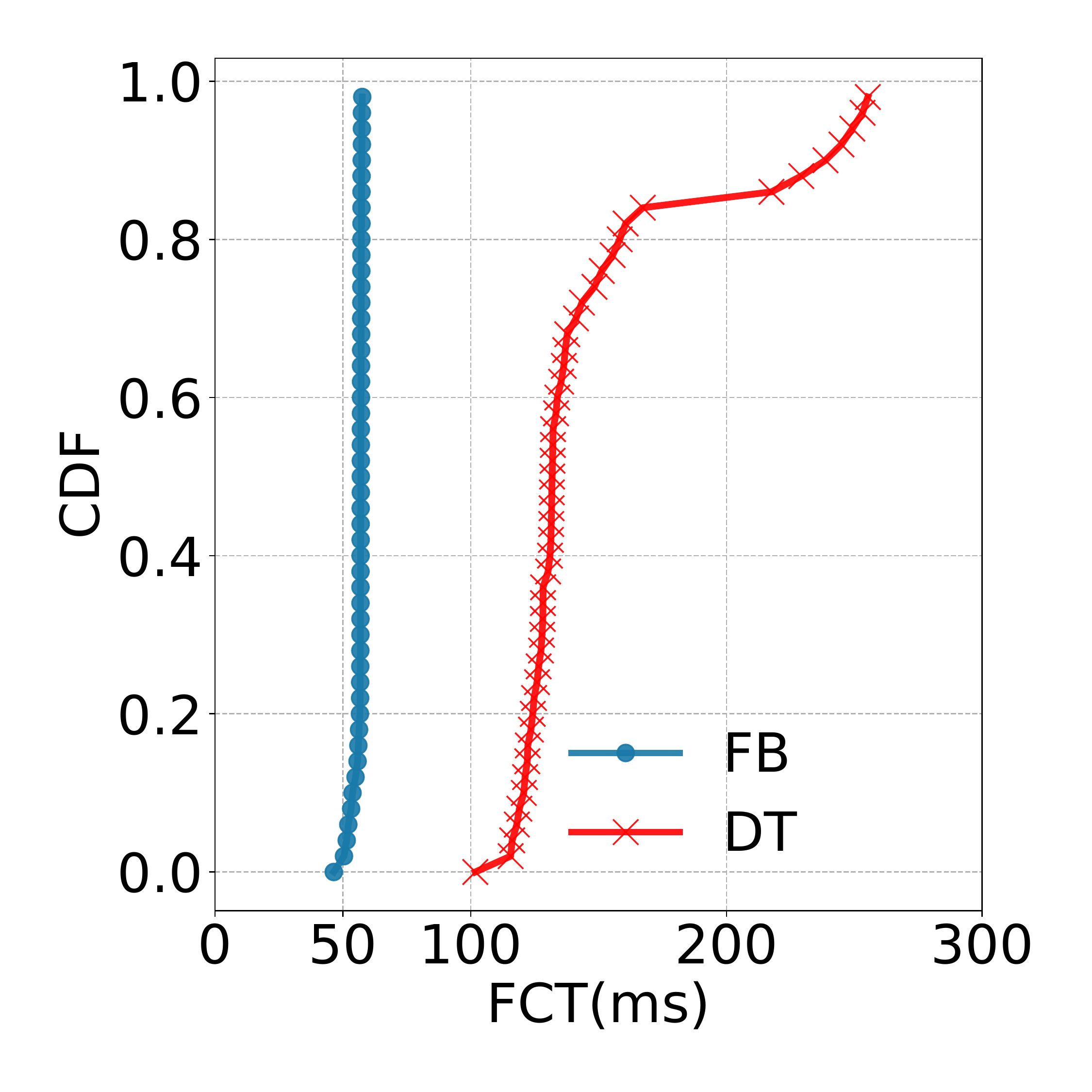}
  \caption[]{}  
  {}
  \label{fig:fct}
 \end{subfigure}
\caption[]{ 
			\textit{\textbf{(a)}} Testbed with Tofino;
			\textit{\textbf{(b)}} \name achieves much lower FCT than
DT~(Default). }
\label{fig:tof2}
\end{figure}

\section{Case study}\label{sec:casestudy}

We implement our hardware design (\S\ref{sec:hwdesign}) on a Barefoot Tofino
Wedge 100BF-32X to verify that it works on real hardware and can improve performance.
Our program includes ~$850$LoC in P4 and uses $6$ stages in the ingress and $3$ stages in the egress pipeline.

Our testbed (Fig.~\ref{fig:testbed}) includes one Tofino switch and
two servers (S1 and S2). S1 connects to the Tofino via port $p1$, while S2 via ports $p2$ and $p3$. 
Each port is mapped to $8$ queues with aggregate bandwidth $80Kbps$ per port.
Among the $8$ queues, $1$ is used for high-priority TCP traffic and $7$ for low-priority UDP traffic.
We configure $\alpha=0.8$ for high-priority traffic and $\alpha=0.6$ for low\footnote{The selelction of $\alpha$ is affected by design choices specific to Tofino and beyond our control. Still, higher $\alpha$ would make DT's performance worse.}.
All queues share the same buffer pool limited to $9000$ cells. 
We did not use all the buffer space provided in a Tofino due to the low-traffic rate.
We generate UDP flows from S1 to S2 on port $p2$.
We also generate $100$ TCP flows $8$KB in size from S2 $p3$ to S1. 
We run the experiment once with DT and once with \name.
Fig.~\ref{fig:fct} presents the CDF of FCT results. 
\name limits the buffer used by low-priority UDP flows, allowing short flows of high-priority to achieve good performance. As a result, \name achieves at least $50$ms smaller FCTs.

\remove{nsg@tofino3:~/maria/plasticine$ ~/tools/p4-graphs-tofino.sh plasticine2.p4
Using SDE /data/bf-sde-8.9.1
Found .manifest
Using SDE_INSTALL /data/bf-sde-8.9.1/install
Your path contains $SDE_INSTALL/bin. Good
WARNING in plasticine2.p4 line 477: Inferred conflicting widths for argument 'port' (8 and 9), using larger width
Generating files in directory /home/nsg/maria/plasticine

TABLE DEPENDENCIES...

INGRESS PIPELINE
['regularPkt']
['syncPkt', 'moveIndex']
['initSync', 'StopRecirculation', 'Npriorities', 'Nports', 'qLengthCP', 'remaining']
['mqlength', 'FindThress']
['compare', 'mthreshold']
['mres', 'Drop']
pipeline ingress requires at least 6 stages

EGRESS PIPELINE
['remove']
['updateIndex']
['qLength']
pipeline egress requires at least 3 stages
}

\section{Related Work}

\name is related but complementary to algorithms operating at the port level (\eg queue management, scheduling) and host level (\eg TCP). 
First, active Queue Management and scheduling algorithms can facilitate preferential treatment of some flows over others, but only if they are mapped to the same output port.
Indeed, AQM pro-actively controls individual queues \eg RED~\cite{red}, Codel~\cite{codel}, PIE~\cite{6602305}, limits the per-flow buffer or bandwidth~\cite{hoeiland2016flowqueue,Ertemalp:2001:UDB:876907.881596,bechtolsheim2003per,Mushtaq:2014:RBM:2740070.2631462} on per-port level.
Second, scheduling techniques \eg pFabric~\cite{alizadeh2013pfabric} and PIAS~\cite{bai2014pias} allow certain flows to be dequeued faster than others of the same port.
Third, congestion control mechanisms, such as ~\cite{alizadeh2011data,saeed2020annulus,dcqcn,gemini,bbr,tcpVegas,timely} reduce the unwanted buffer usage, but cannot detect or react to high overall buffer occupancy, neither can they change the allocated buffer at the device level.

Regarding buffer algorithms more recent works such as EDT~\cite{edt}, FAB~\cite{apostolaki2019fab} and Cisco IB~\cite{cisco9000} empirically recognize bursts and prioritize them. While intuitive, such algorithms cannot consistently (under various loads) absorb bursts as we show in \S\ref{sec:evaluation} and cannot be trivially mapped to multi-queue scenarios. On the contrary, \name offers provable guarantees under various scenarios.
Other buffer management algorithms such as~\cite{830130,1493826,hahne2002dynamic,dynamicPartioning} are not applicable to Call Admission Control (CAC), while pushout-based ones such as~\cite{wei1991optimalPushout},~\cite{thareja1984designPushout} are considered impractical~\cite{survey}.

\section{Conclusion}
In this paper, we demonstrate the inefficiencies of today's most common buffer management algorithm, Dynamic Thresholds, both experimentally and analytically.
We present \name, a novel algorithm that offers provable burst-tolerance guarantees, without sacrificing throughput or
statically allocating buffer.
We show that \name outperforms all other buffer management techniques even when they are combined with DCTCP.
Finally, we show that \name's design is practical by implementing it on a Barefoot Tofino.

\newpage

\bibliographystyle{ACM-Reference-Format}
\bibliography{main}

\newpage

\appendix

\onecolumn

\section{Analysis}
\label{sec:analysis}
\subsection{Assumptions}

The analysis is based on a fluid model where packet (bits) arrivals and departures are assumed to be fluid and deterministic. A switch with arbitrary number of ports with arbitrary number of queues per port is considered. In particular, each port has only one queue per class as defined in \S\ref{sec:background}).

$B$ : Total shared buffer space of switch.

$Q(t)$ : Instantaneous occupied buffer space at time $t$

$\alpha_c$ : A parameter for buffer-management algorithm, for each class.
\subsection{\name}
\name works based on two-levels of hierarchy i.e class and priority. The general notion of class remains same \ie each class is associated with a separate queue at each port. In addition, \name requires the classes to be mapped to priorities. A \emph{Low priority} is a set of classes which share the buffer fairly proportionate to their alpha values. A \emph{High Priority} is simply a set consisting of one class.

\name buffer-management algorithm requires an $\alpha_{c}$ parameter per class. The buffer-allocation is based on threshold calculations per queue. In particular, the threshold of a queue at port $i$, of class $c$ and belonging to a priority $p$ is calculated by \name algorithm as,
\[
\label{plastcine algorithm generalized in words}
Threshold = (alpha) \times(FairShare) \times (Norm.DequeueRate) \times (remaining)
\]
i.e,
\begin{equation}
\label{plasticineAlg}
T_{c}^{i} (t) = \alpha_c \cdot \beta_{p(c)}(t) \cdot \gamma_c^i (t) \cdot (B-Q(t))
\end{equation}
where, $\beta_{p(c)}(t)$ is the inverse of the total number of congested queues of priority $p(c)$ (to which the class $c$ belongs to) at time $t$ \ie $\frac{1}{N_{p}(t)}$ and $\gamma_c^i(t)$ is the normalized dequeue rate (or normalized service rate) of the queue at time $t$.

\noindent Observe that, $\beta_{p(c)}(t)$ remains same for all the priorities belonging to a group and can be expressed as $\beta_{P}(t)$ where $P$ denotes the priority $p(c)$.

\noindent Here after for simplicity $\omega_c^i(t)$ is defined as,

\begin{equation}
\label{omega}
\omega_c^i(t) = \alpha_c \cdot \beta_{P}(t) \cdot \gamma_c^i (t)
\end{equation}

\myitem{Key properties of Omega:}

\begin{equation}
\label{omega General property}
\sum_i \sum_{c \in P} \omega_c^i (t) = \beta_P(t) \cdot \sum_i \sum_{c \in P} \alpha_c \gamma_c^i(t) 
\end{equation}

\noindent If, all the queues of all classes of priority $P$ share same alpha parameters (say $\alpha_P$) and have same normalized dequeue rate at time $t$ (say $\gamma_P$), then Eq.~\ref{omega General property} reduces to,

\[
\sum_i \sum_{c \in P} \omega_c^i (t) = \alpha_P \cdot \gamma_P
\]
Further, if $\gamma_P=1$,
\[
\sum_i \sum_{c \in P} \omega_c^i (t) = \alpha_P
\]

\noindent In general there exists a limit given by,
\begin{equation}
\label{omega max property}
\sum_i \sum_{c \in P} \omega_c^i (t) \le max_{c\in P}(\alpha_c) = \alpha_{max}^P
\end{equation}

We will see later, how these properties enable \name to achieve certain isolation and burst-tolerance guarantees.


\subsection {Steady-State Analysis}

In this subsection, it is assumed that load-conditions remain stable and a steady-state of buffer is achieved. Following this assumption, all the equations in this subsection are expressed without the time variable. Under this state, the queue lengths remain stable at less than or equal to the corresponding threshold. For simplicity, it is assumed that all the queues-lengths are at their respective thresholds. Then the total buffer occupancy can be expressed as,

\[
Q = \sum_i \sum_c Q_c^i
\]
From the assumption that the queue-lengths are equal to their thresholds, using Eq.~\ref{plasticineAlg} and Eq.~\ref{omega},

\[
Q = \sum_i \sum_c \omega_c^i \cdot (B-Q)
\]

\noindent Solving for $Q(t)$ gives,

\begin{equation}
\label{ss.occupancy}
Q = \frac{B \sum_i \sum_c \omega_c^i}{1+\sum_i \sum_c \omega_c^i}
\end{equation}

\noindent where $\omega_c^i$ is given by Eq.~\ref{omega}

\noindent Using, Eq.~\ref{ss.occupancy}, the remaining buffer space $B-Q(t)$ can be expressed as,

\begin{equation}
\label{ss.remaining}
Remaining = \frac{B}{1+\sum_i \sum_c \omega_c^i}
\end{equation}

\noindent Under steady-state, from Eq.~\ref{ss.remaining} and Eq.~\ref{plasticineAlg}, the threshold of a queue at port $i$ and of class $c$ is given by,

\begin{equation}
\label{ss.threshold}
T_c^i = \frac{B\cdot \omega_c^i}{1+\sum_i \sum_c \omega_c^i}
\end{equation}

\myitem{Key properties of Remaining Buffer space:}

\noindent Using the notion of groups, Eq.~\ref{ss.remaining}, can be expanded as,

\[
Remaining = \frac{B}{1+\sum_p\sum_i \sum_{c \in p} \omega_c^i}
\]

\noindent Using the maximum limit of $\omega$ property from Eq.~\ref{omega max property},

\begin{equation}
\label{ss.remaining.limit}
Remaining \ge \frac{B}{1+\sum_p \alpha_{max}^p}
\end{equation}

\noindent This key property shows how the remaining buffer space and buffer-occupancy are bounded.
For example, lets consider, there exists two classes $c_0$ and $c_1$ with alpha parameters $\alpha_0$ and $\alpha_1$, each of which belongs to a separate priority. The remaining buffer space, irrespective of the number of queues and their dequeue rates, Eq.~\ref{ss.remaining.limit} can be expressed as,

\[
Remaining \ge \frac{B}{1+\alpha_0 + \alpha_1}
\]


\subsection{Transient-State Analysis}

Given a steady-state of buffer assuming that all the queue lengths are controlled by a threshold, when traffic to empty queues appear, load conditions change. The new queues increase in length creating changes in the remaining buffer. As a result, the thresholds and queue lengths under go a transient state. Due to the appearance of new queues, $\omega_c^i$ of some of the existing queues get affected due to the changes in $\beta_{c(p)}$ (number of queues belonging to a class $c$ of priority $p$) and $\gamma_c^i$ (normalized dequeue rate). Let $G_e$ denote the set of queues whose $\omega_c^i$ gets affected and $G_{ne}$ denote the set of queues whose $\omega_c^i$ does not get affected. Note that the $\omega_c^i$ values of $G_e$ only reduce. (It is not possible that $\omega_c^i$ increases due the appearance of a new queue). For simplicity lets denote the queue at port $i$ and of class $c$ with ordered pairs as $(i,c)$. The set of ordered pairs of existing queues is denoted as $S_{old}$. The ordered pairs of new queues that trigger transient state are denoted as $S_{new}$. Observe that $S_{old}=C_{ne} \cup C_{e}$.

The arrival rate of traffic at each new queue is denoted by $r$ and the arrival process is fluid and deterministic. At $t=0$,

\begin{equation}
\label{transient t=0 thresh}
T_{c}^i(0)= \frac{\omega_{c}^i\cdot B} {1 + \displaystyle\sum_{\forall(i,c) \in S_{old}} \omega_{c}^i }
\end{equation}

\begin{equation}
\label{transient t=0 queuelength}
Q_{c}^{i}(0)=
\begin{cases}
\frac{\omega_{c}^i\cdot B} {1 + \displaystyle\sum_{\forall(i,c) \in S_{old}} \omega_{c}^i } & \text{, for $\forall(i,c) \in S_{old}$  } \\
0 & \text{ , for $\forall(i,c) \in S_{new}$ }
\end{cases}
\end{equation}

\noindent At $t=0^+$, $\omega^i_c$ of $G_e$ change and remain same for the entire duration of transient state. At the same time, the $\omega^i_c$ of $G_{ne}$ remain unchanged. Hence such changes are assumed to happen and the time variable is dropped for $\omega_c^i$ in the equations.

From Eq.~\ref{plasticineAlg}, the rate of change of thresholds and queue lengths can be expressed as follows,

\begin{equation}
\label{differential tp}
\frac{d T_{c}^i(t)}{dt} = -\omega_{c}^i\cdot\sum_{\forall(i,c) \in S_{old} \cup S_{new}} \frac{dQ^{i}_{c}(t)}{dt}
\end{equation}

\begin{equation}
\label{differential Qip}
\frac{d Q_{c}^{i}(t)}{dt} = 
\begin{cases}
-\gamma_c^i & \text{, if $Q_{c}^{i}(t) > T_{c}(t)$ and $\forall(i,c)\in S_{old}$}\\
max[-\gamma_c^i, min[ \frac{d T_{c}(t)}{dt}, r-\gamma_c^i ]] & \text{, if $Q_{c}^{i}(t) = T_{c}(t)$ and $\forall(i,c)\in S_{old}$}\\
r-\gamma_c^i & \text{, if $Q_{c}^{i}(t) < T_{c}(t)$ and $\forall(i,c)\in S_{new}$}\\
\end{cases}
\end{equation}

\noindent It can be proved by contradiction that $\frac{d T_{c}^i(t)}{dt} \le 0 < r-\gamma_c^i$. Solving Eq.~\ref{differential tp} and Eq.~\ref{differential Qip} for $t=0+$,

\begin{equation}
\label{differential tip expansion}
\left(\frac{d T_{c}^i(t)}{dt}\right)_{(t=0+)} = -\omega_c^i \cdot \left( \sum_{\forall(i,c)\in S_{old}} max[-\gamma_c^i , \frac{d T_{c}(t)}{dt}_{(t=0+)}] \right) - \omega_c^i\cdot \sum_{\forall(i,c)\in S_{new}} (r-\gamma_c^i)
\end{equation}

\noindent Recall that $S_{old}=G_{e}\cup G_{ne}$. All the queues belonging to $G_{e}$, will experience a change in their $\omega_c^i$ values at $t=0^+$ resulting in their queue-lengths greater than threshold. As a result, the rate of change of their queue lengths is their corresponding dequeue rates. Eq.~\ref{differential tip expansion} can then be expanded as,

\begin{equation}
\label{differential tip expansion more}
\begin{aligned}
\left(\frac{d T_{c}^i(t)}{dt}\right)_{(t=0+)} = -\omega_c^i \cdot \left(\sum_{\forall(i,c)\in G_{e}}-\gamma_c^i \right) -\omega_c^i \cdot \left( \sum_{\forall(i,c)\in G_{ne}} max[-\gamma_c^i , \frac{d T_{c}(t)}{dt}_{(t=0+)}] \right) \\
- \omega_c^i\cdot \sum_{\forall(i,c)\in S_{new}} (r-\gamma_c^i)
\end{aligned}
\end{equation}

\noindent From Eq.~\ref{differential tip expansion more}, arrival rate of traffic in new queues i.e $r$ can be expressed as,

\begin{equation}
\label{rate r equation}
 r = \frac{\displaystyle\sum_{\forall(i,c) \in S_{new}\cup G_e} \gamma_c^i}{\displaystyle\sum_{\forall(i,c) \in  S_{new}} 1} - \frac{\frac{d T_{c}^i(t)}{dt}_{(t=0+)} + \omega_c^i\cdot \left( \displaystyle\sum_{\forall(i,c)\in G_{ne}} max[-\gamma_c^i , \frac{d T_{c}(t)}{dt}_{(t=0+)}] \right)}{\omega_c^i \cdot \displaystyle\sum_{\forall(i,c) \in  S_{new}}1}
\end{equation}

\noindent By applying summation across $\forall(i,c)\in G_e$ over Eq.~\ref{differential tip expansion more} (will be seen later how this will be useful), r can be expressed as,

\begin{equation}
\label{rate r equation summation}
 r = \frac{\displaystyle\sum_{\forall(i,c) \in S_{new}\cup G_e} \gamma_c^i}{\displaystyle\sum_{\forall(i,c) \in  S_{new}} 1} - \frac{ (\displaystyle\sum_{i,c \in G_{ne}} \frac{d T_{c}^i(t)}{dt}) + \left( \displaystyle\sum_{\forall(i,c)\in G_{ne}} max[-\gamma_c^i , \frac{d T_{c}(t)}{dt}] \right)\cdot \displaystyle\sum_{\forall(i,c) \in G_{ne}} \omega_c^i}{(\displaystyle\sum_{\forall(i,c) \in G_{ne}} \omega_c^i )\cdot (\displaystyle\sum_{\forall(i,c) \in  S_{new}}1)}
\end{equation}

\noindent Now it can be observed that the value of $r$ influences for all $\forall(i,c)\in G_e$, $\left(\frac{d T_{c}^i(t)}{dt}\right)_{(t=0+)}$. In other words, the value of $r$ influences the total i.e $\sum_{\forall(i,c)\in G_e} \left(\frac{d T_{c}^i(t)}{dt}\right)_{(t=0+)}$ which is the aggregate rate at which thresholds drop for the non affected set of queues i.e $G_{ne}$.

\subsubsection{Case-1}
\label{sec:case-1}
In this case, the arrival rate $r$ is such that, the queues belonging to $G_{ne}$ are able to reduce in length exactly tracking the changes in their thresholds. As a result their queue-lengths remain equal to the threshold throughout the transient state i.e,

\begin{equation}
\label{case1 assumption1}
\left(\frac{d T_{c}^i(t)}{dt}\right)_{(t=0^{+})}\ge-\gamma_c^i
\end{equation}
leading to,
\begin{equation}
\label{case1 assumption2}
\sum_{\forall(i,c)\in G_{ne}} \left(\frac{d T_{c}^i(t)}{dt}\right)_{(t=0+)} \ge \sum_{\forall(i,c)\in G_{ne}} -\gamma_c^i
\end{equation}
\newline
Using Eq.~\ref{case1 assumption1} and Eq.~\ref{case1 assumption2} in Eq.~\ref{rate r equation summation}, the condition on $r$ can be expressed as,
\newline
\begin{equation}
\label{case1}
 r \le \frac{\displaystyle\sum_{\forall(i,c) \in S_{new}\cup G_e} \gamma_c^i}{\displaystyle\sum_{\forall(i,c) \in  S_{new}} 1} + \left(\sum_{\forall(i,c)\in G_{ne}}^* \gamma_c^i\right)\cdot \frac{ 1 + \displaystyle\sum_{\forall(i,c) \in G_{ne}} \omega_c^i}{(\displaystyle\sum_{\forall(i,c) \in G_{ne}}^* \omega_c^i )\cdot (\displaystyle\sum_{\forall(i,c) \in  S_{new}}1)}
\end{equation}
\newline

\noindent Note that in Eq.~\ref{rate r equation summation}, we deliberately apply summation over $\forall(i,c)\in G_{ne}$ which can be a null set. If $G_{ne}=\phi$, by applying summation over $\forall(i,c)\in G_{e}$ in Eq.~\ref{differential tip expansion more}, $r$ condition can be expressed as, 

\begin{equation}
\label{case1 ifnot}
 r \le \frac{\displaystyle\sum_{\forall(i,c) \in S_{new}} \gamma_c^i}{\displaystyle\sum_{\forall(i,c) \in  S_{new}} 1} + \left(\sum_{\forall(i,c)\in G_{e}} \gamma_c^i\right)\cdot \frac{ 1 + \displaystyle\sum_{\forall(i,c) \in G_{e}} \omega_c^i}{(\displaystyle\sum_{\forall(i,c) \in G_{e}} \omega_c^i )\cdot (\displaystyle\sum_{\forall(i,c) \in  S_{new}}1)}
\end{equation}
\newline

For generalization, observe the ``*'' over the summation terms in Eq.~\ref{case1}. Here after, the convention follows that, where ever ``*'' appears, it means that, the summation is deliberate and can be interchanged between $\forall(i,c)\in G_{ne}$ and $\forall(i,c)\in G_{e}$ if $G_{ne}=\phi$. All the other summations have usual meaning.

\noindent Substituting Eq.\ref{case1 assumption1} and Eq.\ref{case1 assumption2} in Eq.~\ref{differential tip expansion more} and using the result in Eq.~\ref{differential Qip} gives,

\begin{equation}
\label{differential tip solved case1.1}
\left(\frac{d T_{c}(t)}{dt}\right)_{(t=0^{+})} = \frac{\displaystyle - \omega_{c}^i\cdot \left( \sum_{\forall(i,c)\in G_e} -\gamma_c^i  + \sum_{\forall(i,c)\in S_{new}} (r-\gamma_c^i) \right)} {1+\displaystyle\sum_{\forall(i,c)\in G_{ne}} \omega_{c}^i}
\end{equation}

\begin{equation}
\label{differential Qip solved case1.1}
\left(\frac{d Q_{c}^{i}(t)}{dt}\right)_{(t=0^{+})} = 
\begin{cases}
\frac{\displaystyle - \omega_{c}^i\cdot \left( \sum_{\forall(i,c)\in G_e} -\gamma_c^i  + \sum_{\forall(i,c)\in S_{new}} (r-\gamma_c^i) \right)} {1+\displaystyle\sum_{\forall(i,c)\in G_{ne}} \omega_{c}^i} & \text{, for $\forall(i,c)\in G_{ne}$}\\
-\gamma_c^i & \text{, for $\forall(i,c)\in G_e$}\\
r-\gamma_c^i & \text{, for $\forall(i,c)\in S_{new}$}\\
\end{cases}
\end{equation}

\noindent These differential equations will be valid as long as $Q^{i}_{c}(t) = T_{c}^i(t)$ for $\forall(i,c)\in G_{ne}$ $\&\&$ $Q^{i}_{c}(t) \ge T_{c}^i(t)$ for $\forall(i,c)\in G_{e}$ $\&\&$ $Q^{i}_{c}(t) < T_{c}^i(t)$ for newly created queues i.e $\forall(i,c)\in S_{new}$. Solving these equation, using the initial conditions, Eq.~\ref{transient t=0 thresh} and Eq.~\ref{transient t=0 queuelength} leads to,

\begin{equation}
\label{tp solved case1}
T_{c}^i(t)= \frac{\omega_{c}^i\cdot B} {1 + \displaystyle\sum_{\forall(i,c) \in S_{old}} \omega_{c}^i } - \frac{\displaystyle \omega_{c}^i\cdot \left( \sum_{\forall(i,c)\in G_e} -\gamma_c^i  + \sum_{\forall(i,c)\in S_{new}} (r-\gamma_c^i) \right)\cdot t} {1+\displaystyle\sum_{\forall(i,c)\in G_{ne}} \omega_{c}^i}
\end{equation}

\begin{equation}
\label{qip solved case1}
Q_{c}^{i}(t)=
\begin{cases}
\frac{\displaystyle\omega_{c}^i\cdot B} {1 + \displaystyle\sum_{\forall(i,c) \in S_{old}} \omega_{c}^i } - \frac{\displaystyle \omega_{c}^i\cdot t\cdot ( \sum_{\forall(i,c)\in G_e} -\gamma_c^i +\sum_{\forall(i,c)\in S_{new}} (r-\gamma_c^i) )} {1+\displaystyle\sum_{\forall(i,c)\in G_{ne}} \omega_{c}^i} & \text{, for $\forall (i,c)\in G_{ne}$  } \\
\frac{\displaystyle\omega_{c}^i\cdot B} {1 + \displaystyle\sum_{\forall(i,c) \in S_{old}} \omega_{c}^i } - \gamma_c^i \cdot t & \text{,for $\forall (i,c)\in G_e$}\\
(r-\gamma_c^i)\cdot t & \text{, for $\forall (i,c)\in S_{new}$ }
\end{cases}
\end{equation}

\noindent As we can observe from Eq.~\ref{tp solved case1} and Eq.~\ref{qip solved case1}, the new queues will grow in length without dropping packets upto a time say $t1_c^i$ when the threshold equals the queue length. It is considered that $t1_c^i$ denotes the time at which a new queue of class $c$ at port $i$ first touches the threshold. The transient state continues after $t1_c^i$ until all the queues achieve a steady state occupancy. By equating Eq.~\ref{tp solved case1} and Eq.~\ref{qip solved case1} for the case of $\forall (i,c)\in S_{new}$, $t1_c^i$ can be obtained as,

\begin{equation}
\label{t1 case1}
t1_c^i = \frac{\omega_{c}^i\cdot B \cdot (1+\displaystyle\sum_{\forall(i,c)\in G_{ne}} \omega_{c}^i ) } {(1 + \displaystyle\sum_{\forall(i,c) \in S_{old}} \omega_{c}^i )\cdot ( (r-\gamma_c^i)\cdot (1+\displaystyle\sum_{\forall(i,c)\in G_{ne}} \omega_{c}^i ) + \displaystyle \omega_{c}^i\cdot ( \sum_{\forall(i,c)\in G_e} -\gamma_c^i  + \sum_{\forall(i,c)\in S_{new}} (r-\gamma_c^i) ) ) }
\end{equation}

\noindent In order to offer guarantees, it is absolutely required that either $\gamma_c^i$ is constant. The reason being that there is a dependency between $\gamma_c^i$ and the number of queues of the same port using buffer, a dependency which is fundamentally impossible to evade unless $\gamma_c^i$ is constant. As a result of this assumption, $G_e=\phi$ and $S_{old}=G_{ne}$ and Eq.~\ref{t1 case1} reduces to,

\begin{equation}
\label{t1 case1 simp}
t1_c^i = \\
\frac{\alpha_H \cdot \frac{1}{N_{p(c)}}\cdot\gamma_c^i \cdot B } { (r-\gamma_c^i) \cdot (1+\displaystyle\sum_{\forall(i,c)\in S_{old}} \omega_{c}^i  + \displaystyle \omega_{c}^i\cdot\sum_{\forall(i,c)\in S_{new}} 1 ) }\\
\end{equation}

\noindent We can further simplify for a case where load variations occur for \emph{High Priority} whose maximum $\alpha$ value is $\alpha_H$ and the existing \emph{Low Priority} in the queues have a maximum $\alpha$ value of $\alpha_L$. We can then guarantee that for an arrival rate $r$ that satisfies \emph{Case-1} will experience zero drops \ie no transient drops if its duration $t$ satisfies the following condition: 
\begin{equation}
\label{t1 case1 simplified}
t1_c^i = \\
\frac{\alpha_H \cdot \frac{1}{N_{p(c)}}\cdot\gamma_c^i \cdot B } {(r-\gamma_c^i)\cdot \left(1+\alpha_L + \displaystyle \alpha_H \cdot \frac{1}{N_{p(c)}}\cdot\gamma_c^i\cdot \sum_{\forall(i,c)\in S_{new}} 1\right)}\\
\end{equation}

\noindent Observe that Eq.~\ref{t1 case1 simplified} is independent of number of queues of \emph{Low Priority} and hence it can be said that \emph{High Priority} isolation can be guaranteed.

\subsubsection{Case-2}
In this case, the arrival rate $r$ is such that, the queues belonging to $G_{ne}$ are unable to reduce in length in accordance with the changes in their thresholds. As a result their queue-lengths remain greater than the threshold throughout the transient state i.e,

\begin{equation}
\label{case2 assumption1}
\left(\frac{d T_{c}^i(t)}{dt}\right)_{(t=0^{+})} < -\gamma_c^i
\end{equation}
leading to,
\begin{equation}
\label{case2 assumption2}
\sum_{\forall(i,c)\in G_{ne}} \left(\frac{d T_{c}^i(t)}{dt}\right)_{(t=0+)} < \sum_{\forall(i,c)\in G_{ne}} -\gamma_c^i
\end{equation}
\newline
Using Eq.~\ref{case2 assumption1} and Eq.~\ref{case2 assumption2} in Eq.~\ref{rate r equation summation}, the condition on $r$ can be expressed as,
\newline
\begin{equation}
\label{case2}
 r > \frac{\displaystyle\sum_{\forall(i,c) \in S_{new}\cup G_e} \gamma_c^i}{\displaystyle\sum_{\forall(i,c) \in  S_{new}} 1} + \left(\sum_{\forall(i,c)\in G_{ne}}^* \gamma_c^i\right)\cdot \frac{ 1 + \displaystyle\sum_{\forall(i,c) \in G_{ne}} \omega_c^i}{(\displaystyle\sum_{\forall(i,c) \in G_{ne}}^* \omega_c^i )\cdot (\displaystyle\sum_{\forall(i,c) \in  S_{new}}1)}
\end{equation}
\newline

\noindent if $G_{ne}=\phi$, then $r$ the above condition is expressed as,

\begin{equation}
\label{case2 ifnot}
 r > \frac{\displaystyle\sum_{\forall(i,c) \in S_{new}} \gamma_c^i}{\displaystyle\sum_{\forall(i,c) \in  S_{new}} 1} + \left(\sum_{\forall(i,c)\in G_{e}} \gamma_c^i\right)\cdot \frac{ 1 + \displaystyle\sum_{\forall(i,c) \in G_{e}} \omega_c^i}{(\displaystyle\sum_{\forall(i,c) \in G_{e}} \omega_c^i )\cdot (\displaystyle\sum_{\forall(i,c) \in  S_{new}}1)}
\end{equation}
\newline

\noindent Following similar procedure as in \emph{Case-1}, the equations for \emph{Case-2} can be easily determined. Finally, the time $t1_c^i$ at which one of the queues that belong to $S_{new}$ touches it's threshold can be expressed as,

\begin{equation}
\label{t1 case2}
t1_c^i = \frac{\displaystyle\omega_{c}^i\cdot B  } {\left(1 + \displaystyle\sum_{\forall(i,c) \in S_{old}} \omega_{c}^i \right)\cdot \left( (r-\gamma_c^i) + \displaystyle \omega_{c}^i\cdot \left( \sum_{\forall(i,c)\in S_{old}} -\gamma_c^i  + \sum_{\forall(i,c)\in S_{new}} (r-\gamma_c^i) \right) \right) }
\end{equation}

\noindent Further based on $\omega$ properties and observing that $\sum_{\forall(i,c)\in S_{old}} -\gamma_c^i = (-)$\emph{Number of congested ports of $S_{old}$} say $-NUM$.

\begin{equation}
\label{t1 case2 simplified}
t1_c^i = \frac{\alpha_H \cdot \frac{1}{N_{p(c)}}\cdot\gamma_c^i\cdot B  } {\left(1 + \alpha_L \right)\cdot \left( (r-\gamma_c^i) + \displaystyle \alpha_H \cdot \frac{1}{N_{p(c)}}\cdot\gamma_c^i\cdot \left( -NUM  + \sum_{\forall(i,c)\in S_{new}} (r-\gamma_c^i) \right) \right) }
\end{equation}

\noindent Notice that the presence of NUM in Eq.~\ref{t1 case2 simplified}, is a dependency on the number of congested ports of \emph{Low Priority}. However, \emph{NUM} only creates a positive effect on $t1_c^i$ \ie greater the $NUM$ greater is $t1_c^i$. On the other hand, Eq.~\ref{t1 case2 simplified} is independent of negative dependencies as was in the traditional algorithm DT.
\newline

\subsection{How it all relates to Burst-Tolerance}
\label{sec:guarantees}
First, if only arrival rate $r$ is known and an operator wishes to guarantee zero transient losses, from Eq.~\ref{case1}, assuming that load variation occur on an empty port, $\alpha_L$ is upper bounded by,

\begin{equation}
\label{Zerotransient}
\alpha_L \le \frac{1}{r-(NUM+1)}
\end{equation}

\noindent where $n$ is the number of congested ports. Observe that the worst case for \name is when only a single queue is congested at $t=0$ when the burst arrives. $\alpha_L$ can be easily determined by using good enough estimate on the number of congested ports $NUM$. When the actual number of congested ports are $<NUM$, \name cannot guarantee burst absorption. On the other hand, \name guarantees burst absorption for any number of congested ports $>NUM$. In practice, this is a desirable property as one could use a conservatively low number for $n$ to determine $\alpha_L$ to guarantee burst absorption. If an operator still wants to achieve $100\%$ burst absorption, $NUM=1$ must be used as follows,

\[
\label{Zerotransient}
\alpha_L \le \frac{1}{r-2}
\]

Denote Burst-Tolerance for a queue of class $p$ at port $i$ as $Burst_c^i$ can be defined as

\begin{equation}
Burst_c^i= r\cdot t1_c^i
\end{equation}

\noindent where $r$ is the arrival rate of traffic.

\noindent Then, the maximum burst that can pass without experiencing drops is given by $Burst_c^i$. Say an operator specifies $Burst_c^i$ i.e $r$ and $t1_c^i$ to be guaranteed to pass at all times. How can $\alpha_c$ be optimized?

Given an arrival rate $r$ and of duration $t$ on a single queue, at a given state of buffer, we are interested in providing a guarantee such that the burst is successfully absorbed. Hence we consider worst case scenarios to derive bounds on $\alpha_L$ and $\alpha_H$, where $\alpha_L$ is the maximum value for \emph{Low Priority} and $\alpha_H$ is the maximum for \emph{High Priority}. 
Additionally for simplicity, it is assumed that a burst happens on an empty port leading to $\gamma_c^i=1$ and $G_{e}=\phi$.

\begin{figure*}[!h]
\centering
\begin{subfigure}{0.45\linewidth}
\centering
\includegraphics[width=1\linewidth]{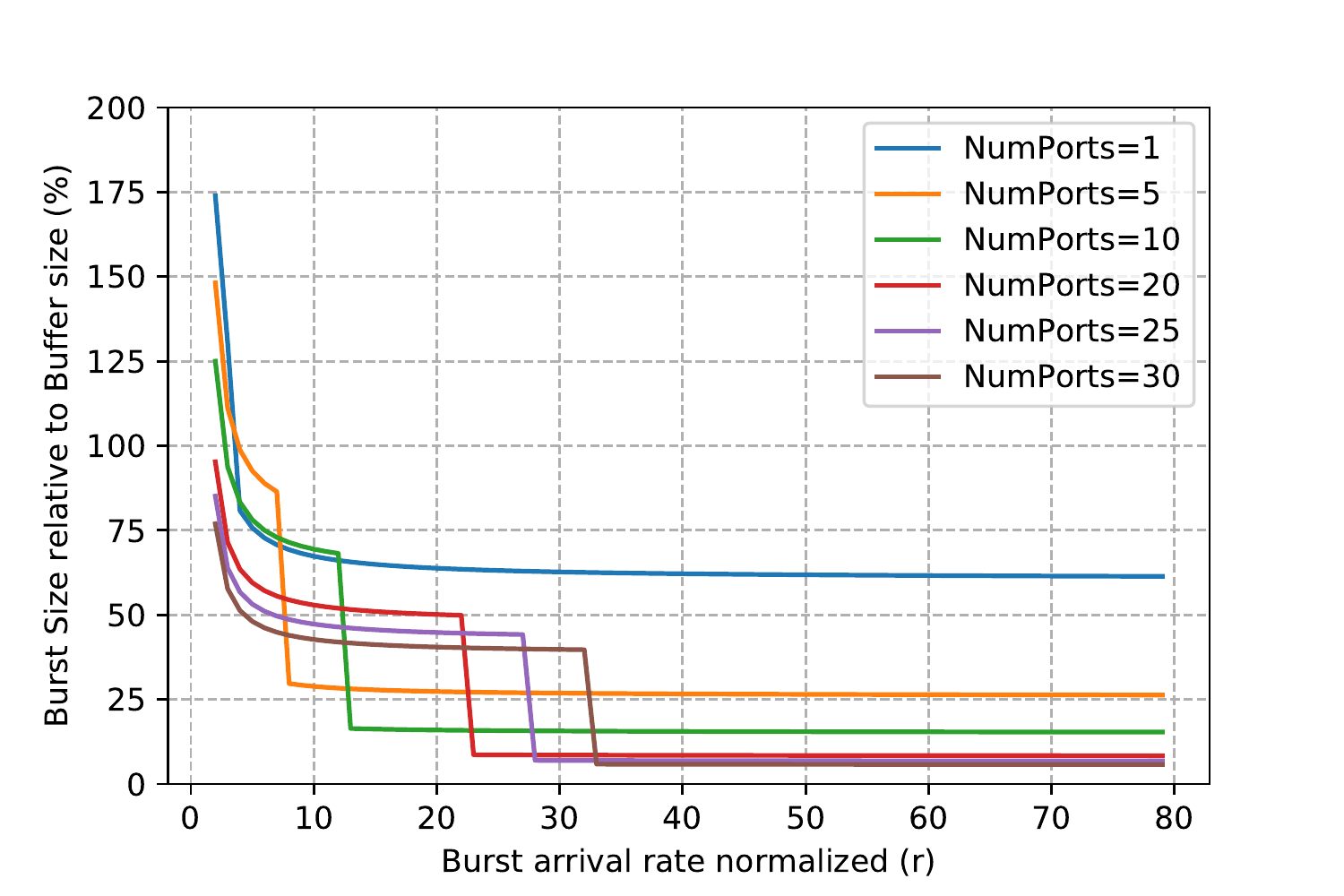}
\caption{Dynamic Thresholds (Single Queue)}
\vspace{2mm}
\label{fig:DtBurstAnalysis}
\end{subfigure}
\begin{subfigure}{0.45\linewidth}
\centering
\includegraphics[width=1\linewidth]{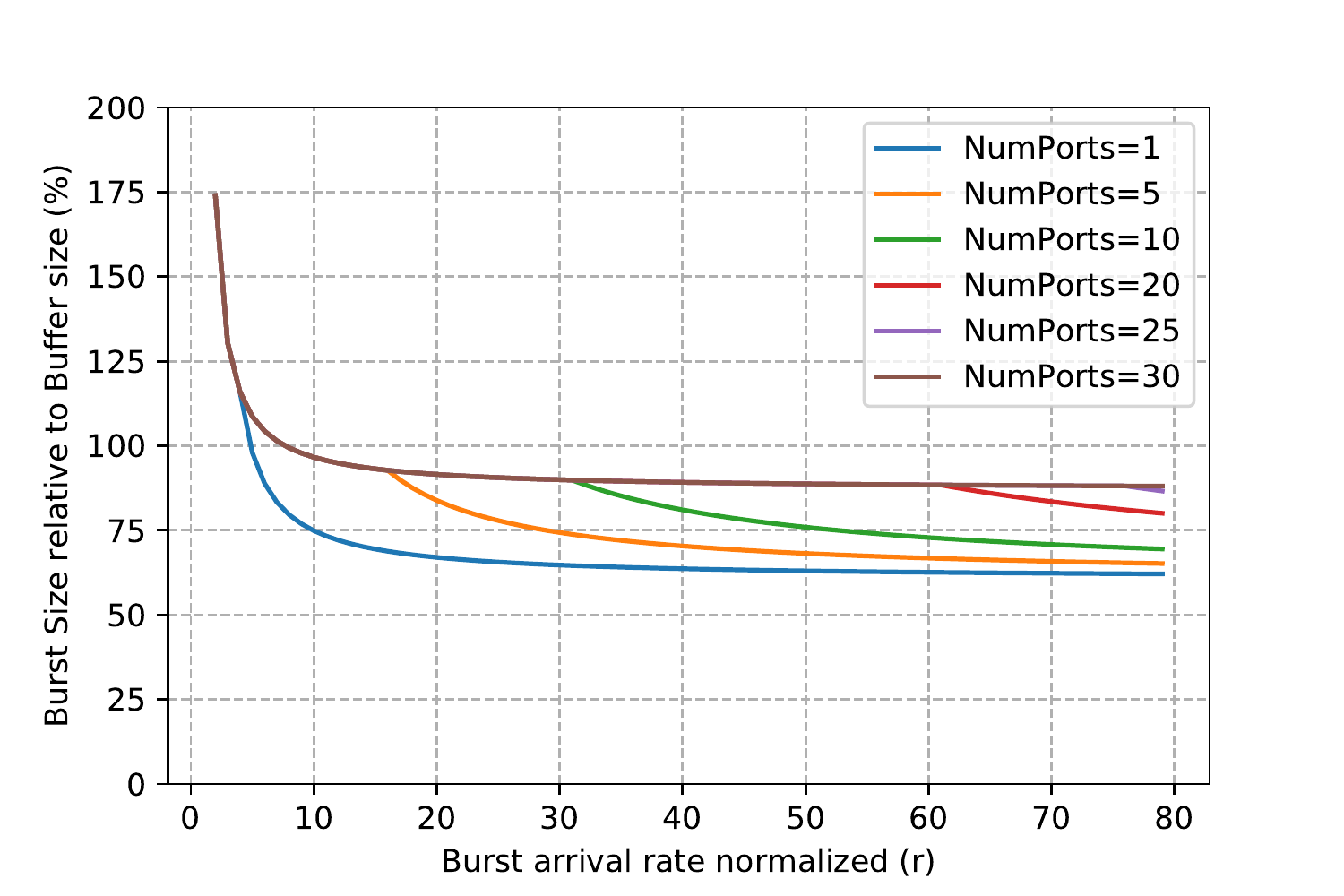}
\caption{\name (Single Queue)}
\vspace{2mm}
\label{fig:PcBurstAnalysis}
\end{subfigure}
\begin{subfigure}{0.45\linewidth}
\centering
\includegraphics[trim=0 5 0 0,clip,width=1\linewidth]{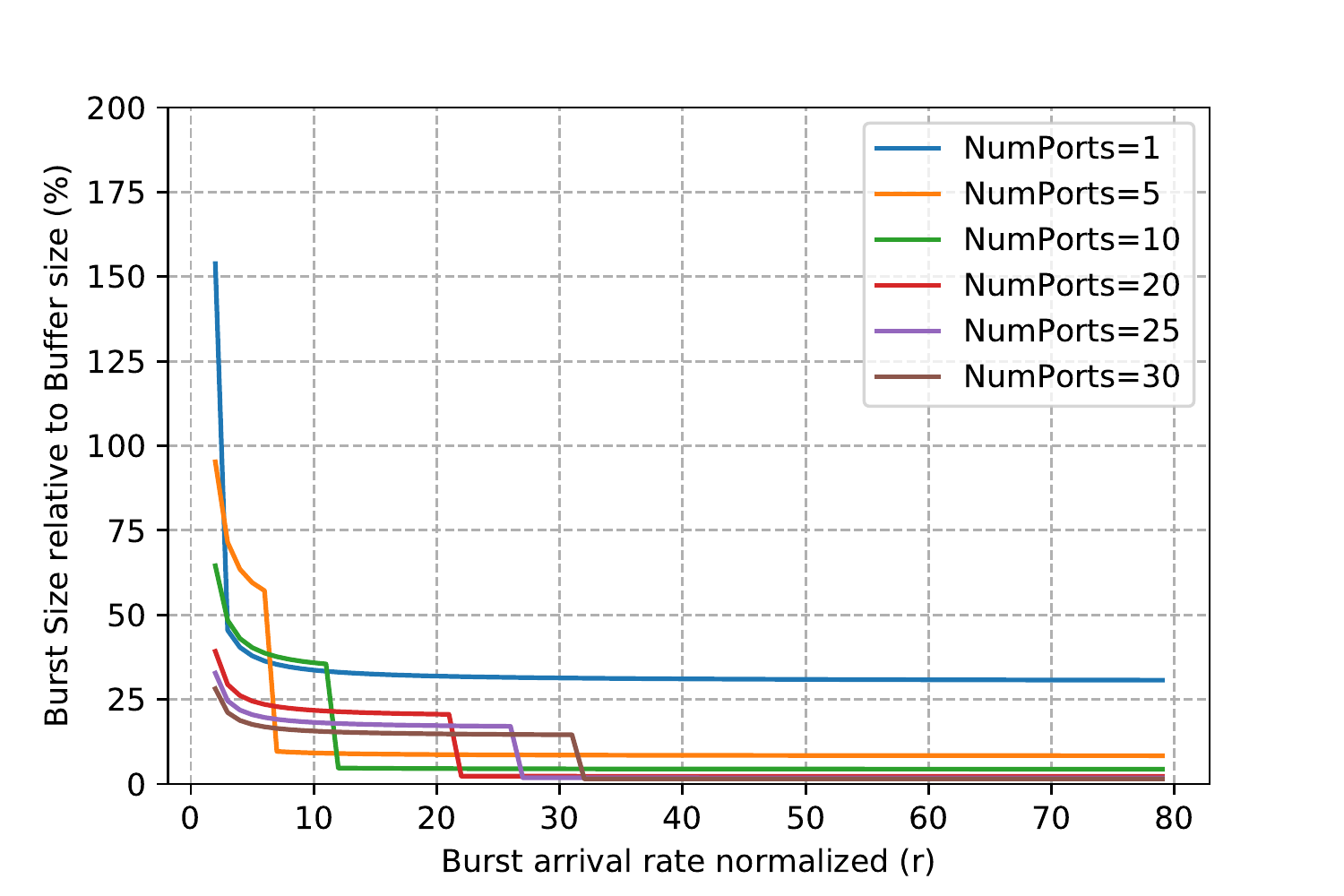}
\caption{Dynamic Thresholds (4 Queues)}
\label{fig:DtBurstAnalysisMul}
\end{subfigure}
\begin{subfigure}{0.45\linewidth}
\centering
\includegraphics[trim=0 5 0 0,clip,width=1\linewidth]{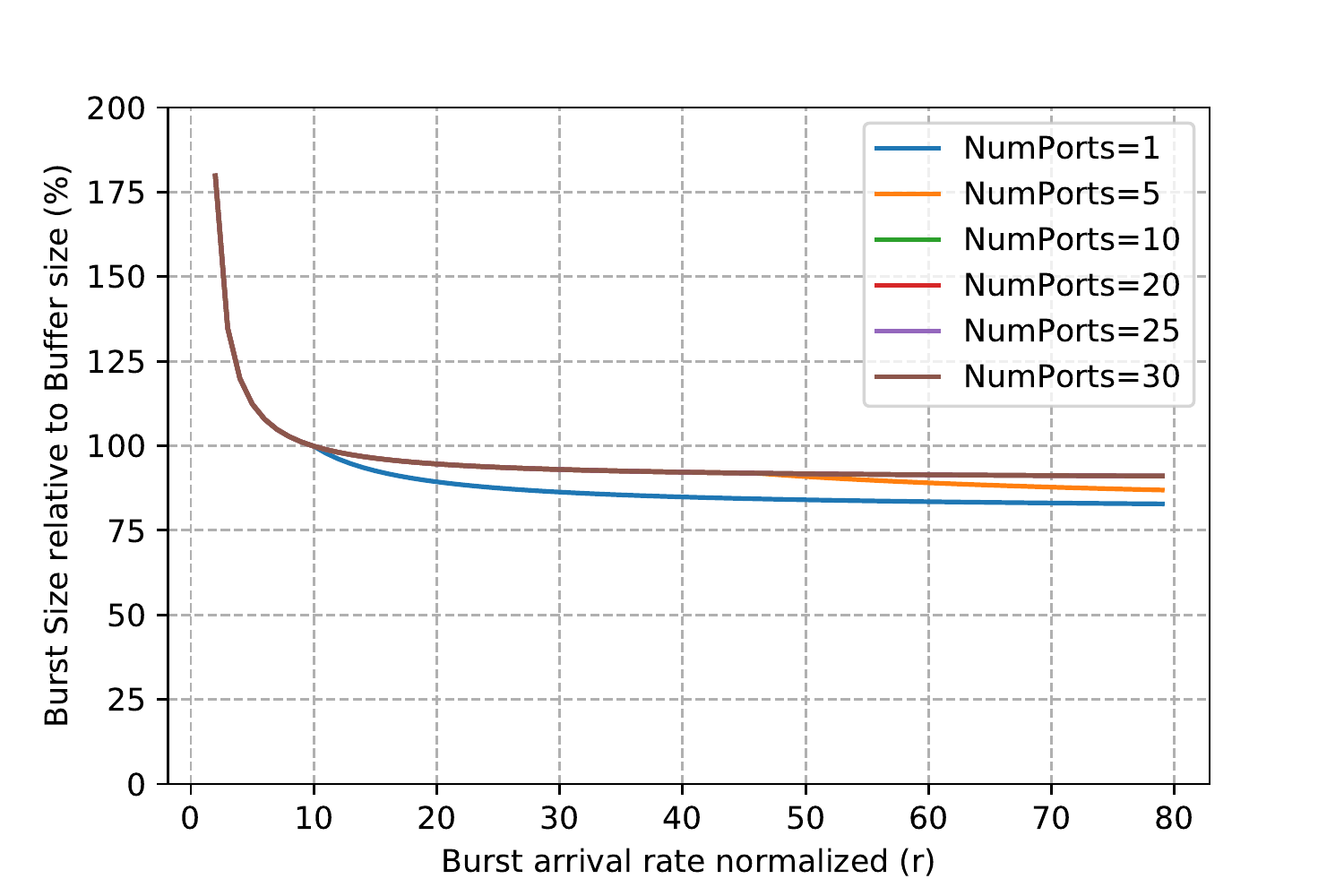}
\caption{\name (4 Queues)}
\label{fig:PcBurstAnalysisMul}
\end{subfigure}
\caption{Comparison of the burst absorption capabilities of \name and DT showing how \name's performance guarantees remain unaffected by the state of buffer. Parameters used are same as in (\S\ref{sec:evaluation})}
\label{fig:analysisBurst}
\end{figure*}

Considering a worst case arrival rate $r$ that only satisfies \emph{Case-2}, we are interested in the buffer required and made available by buffer management without drops \ie $(r-1)\cdot t \le (r-1)\cdot t1^c_p$. Using Eq.~\ref{t1 case2 simplified} and letting $NUM=1$ to consider worst case,

\[
(r-1)\cdot t \le \frac{(r-1) \cdot \alpha_H \cdot B  } {\left(1 + \alpha_L \right)\cdot \left( (r-1) + \displaystyle \alpha_H \cdot \left( -1  +  (r-1) \right) \right) }
\]
For an arbitrarily large value of $\alpha_H$, there exists a limit such that,

\begin{equation}
\label{alphaL}
\alpha_L \le \frac{B}{(r-2)\cdot t} -1
\end{equation}

Similarly, using Eq.~\ref{t1 case2 simplified},

\begin{equation}
\label{alpha_H}
\alpha_H > \frac{1}{\frac{B}{(r-1)\cdot t \cdot (1+\alpha_L)}-\frac{r-2}{r-1}}
\vspace{5mm}
\end{equation}

Futher generalizing from the properties of Omega, for a burst that occurs on a queue at port $i$ and of class $c$, for \emph{Case-1} and \emph{Case-2}, from Eq.~\ref{t1 case1} and Eq.~\ref{case2}, the conditions in terms of $\omega_c^i$ can be expressed as,

\begin{equation}
\label{alpha L}
\alpha_{L} 
\begin{cases}
\le \frac{1}{ \displaystyle\left(\frac{ \displaystyle\sum_{\forall(i,c)\in S_{new}} 1} {\displaystyle\sum_{\forall(i,c)\in G_{ne}}^* \gamma_c^i}\right)\cdot\left( r- \left(\frac{\displaystyle\sum_{\forall(i,c)\in S_{new}\cup G_{e}} \gamma_c^i} {\displaystyle\sum_{\forall(i,c)\in S_{new}} 1}\right) \right) -1\cdot!( G_{ne==\phi}) } & \text{\emph{Case-1}}\\
\\
> \frac{1}{ \displaystyle\left(\frac{ \displaystyle\sum_{\forall(i,c)\in S_{new}} 1} {\displaystyle\sum_{\forall(i,c)\in G_{ne}}^* \gamma_c^i}\right)\cdot\left( r- \left(\frac{\displaystyle\sum_{\forall(i,c)\in S_{new}\cup G_{e}} \gamma_c^i} {\displaystyle\sum_{\forall(i,c)\in S_{new}} 1}\right) \right) -1\cdot!( G_{ne==\phi}) } & \text{\emph{Case-2}}
\end{cases}
\end{equation}

Let $\alpha_H$ denote the alpha parameter for the new queues. For a burst with arrival rate $r$ upto time $t$ to pass, it is required that $t<=t1_c^i$.
Then, whether $\alpha_L$ is determined based on the above guideline or chosen based on steady-state allocations, $\alpha_H$ can be expressed as,

\begin{equation}
\label{alpha H} 
\alpha_{H} \ge \frac{1}{\gamma_c^i\cdot\beta^i_{p(c)}} \cdot \frac{  t\cdot (r-\gamma_c^i)\cdot(1+\alpha_L)} {\displaystyle  B - t\cdot \left( \sum_{\forall(i,c)\in G_e} -\gamma_c^i  + \sum_{\forall(i,c)\in S_{new}} (r-\gamma_c^i) \right) }
\end{equation}

Note that, in Eq.~\ref{alpha H}, when the denominator is less than or equal to 0, $\alpha_H$ has no meaning, indicating an impossibility.

\vspace{5mm}
Finally, \name also scales to multiple priority levels. Our analysis considers a generalized model with arbitrary number of priorities. We observe that, \name's allocation scheme regardless of number of queues of each priority using the buffer can always guarantee performance for the highest priority. For instance, let $\alpha_x=\alpha_1+\alpha_2+\alpha_3......+\alpha_n$ where $\alpha_n$ is the maximum $\alpha$ of $n_{th}$ priority, then in order to guarantee a burst of incoming rate $r$ and duration $t$ to pass at all times, $\alpha_H$ (maximum $\alpha$ across the highest priority) can still be derived by simply replacing $\alpha_L$ with $\alpha_x$ in Eq.~\ref{alpha_H}.

One could use the above analysis of transient state and generate analytical plots as shown in Fig.~\ref{fig:PcBurstAnalysis}, Fig.~\ref{fig:PcBurstAnalysisMul} (Showing \name), Fig.~\ref{fig:DtBurstAnalysis}, Fig.~\ref{fig:DtBurstAnalysisMul} (showing DT). The figures show, for a given $\alpha$ parameter setting, the variation of burst absorption (y axis) when the arrival rate changes. Further different lines correspond to different buffer states (with different number of pre occupied low priority queues). A buffer size ($B$) of 1MB, link capacity of 1Gbps, $\alpha_L=0.5$ and $\alpha_H=20$ are used in the equations. Notice that this setting is same as in (\S\ref{sec:evaluation}) enabling a comparison against analysis and simulation results.

 We notice that, DT neither has an upper bound nor a lower bound. On the other hand, the strategic allocation of \name allows for a lower bound (corresponding to a buffer state with single queue) and as the arrival rate increases the performance of various buffer states tends towards the lower bound. As a result of such bound in burst absorption, an operator could easily guarantee the absorption of a burst (corresponding to the lower bound).

\label{lastpage}
\end{document}